\documentclass[12pt]{article}
\usepackage[margin=1in]{geometry}
\usepackage{amsmath}
\usepackage{amssymb,amsthm}
\usepackage[authoryear]{natbib}
\usepackage{titletoc}
\usepackage[dvipsnames]{xcolor}
\usepackage{enumitem}
\usepackage{rotating}
\usepackage{lscape}
\usepackage{float}
\usepackage{graphicx}
\usepackage{mathtools}
\usepackage{accents}

\usepackage{microtype} 

\usepackage{accents}
\usepackage{datetime}

\newdateformat{monthyeardate}{%
  \monthname[\THEMONTH], \THEYEAR}

\usepackage{bm}
\usepackage[multiple]{footmisc}

\usepackage[hidelinks]{hyperref}

\newtheorem{theorem}{Theorem}
\newtheorem{proposition}{Proposition}

\newtheorem{lemma}{Lemma}

\newtheorem{definition}{Definition}
\newtheorem{thm}{Theorem}
\newtheorem{example}{Example}

\theoremstyle{definition}
\newtheorem{remark}[thm]{Remark}

\usepackage{chngcntr}
\usepackage{apptools}
\AtAppendix{\counterwithin{theorem}{section}}

\newtheoremstyle{named}{}{}{\itshape}{}{\bfseries}{.}{.5em}{#1}
\theoremstyle{named}

\providecommand{\abs}[1]{\left\lvert#1 \right\rvert}
\providecommand{\norm}[1]{\left\lVert#1 \right\rVert}
\newcommand{\normo}[1]{{\left\vert\kern-0.25ex\left\vert\kern-0.25ex\left\vert #1 
    \right\vert\kern-0.25ex\right\vert\kern-0.25ex\right\vert}}
\providecommand{\pref}{\succcurlyeq}

\renewcommand{\Re}{\operatorname{\mathbb{R}}}

\DeclareMathOperator{\cl}{cl}
\renewcommand{\P}{\mathcal P}
\providecommand{\W}{\mathcal W}

\providecommand{\U}{\mathcal U}

\providecommand{\Na}{\mathbb N}
\newcommand\xqed[1]{%
  \leavevmode\unskip\penalty9999 \hbox{}\nobreak\hfill
  \quad\hbox{#1}}
\newcommand\myqed{\xqed{$\blacksquare$}}

\providecommand{\F}{\phi}
\providecommand{\G}{\psi}

\providecommand{\A}{\mathcal A}

\DeclareMathOperator{\co}{co}

\DeclareMathOperator{\dom}{dom}

\usepackage{accents}
\newcommand{\ubar}[1]{\underaccent{\bar}{#1}}
\providecommand{\xh}{\bar x}
\providecommand{\xl}{\ubar x}

\makeatletter
\newcommand{\longdash}[1][2em]{
  \makebox[#1]{$\m@th\smash-\mkern-7mu\cleaders\hbox{$\mkern-2mu\smash-\mkern-2mu$}\hfill\mkern-7mu\smash-$}}
\makeatother
\newcommand{\omitskip}{\kern-\arraycolsep}
\newcommand{\llongdash}[1][2em]{\longdash[#1]\omitskip}
\newcommand{\rlongdash}[1][2em]{\omitskip\longdash[#1]}

\DeclareMathOperator{\cone}{cone}
\providecommand{\indic}{\bm 1}

\providecommand{\V}{\mathcal V}

\providecommand{\abold}{\bm a}

\newcommand\restr[2]{{
  \left.\kern-\nulldelimiterspace 
  #1\! 
  \,\vphantom{\big|} 
  \right|_{#2} 
  }}

\newenvironment{examplec}[2][Example]{\begin{trivlist}
\item[\hskip \labelsep {\bfseries #1}\hskip \labelsep {\hspace{-0.2em}\bfseries #2.}]}{\end{trivlist}}

\usepackage{fullpage}

\usepackage{natbib}

\usepackage{mathrsfs}
\providecommand{\w}{\bm{\mathrm{ w}}}

\begin{document}

\title{On the stability of utilitarian aggregation} 

\author{Leandro Nascimento\thanks{Universidade de Bras{\' i}lia; lgnascimento@unb.br.}}

\date{\monthyeardate\today}

\maketitle


\begin{abstract} 

In the context of aggregating von Neumann-Morgenstern utilities, we show that bounded violations of the Pareto conditions characterize aggregation rules that are approximately utilitarian. When a single utility function is intended to represent the preference judgments of a group of individuals and the Pareto principles are nearly satisfied, we prove that its distance from a weighted sum of individual cardinal utilities does not exceed half of the positive parameter that differentiates our weaker versions of the Pareto conditions from their conventional forms. This result suggests the stability of \citeauthor{harsanyi1955}'s  (\citeyear{harsanyi1955}) aggregation theorem, in that small deviations from the Pareto principles lead to aggregation rules that remain close to utilitarian aggregation.

 \bigskip



\end{abstract}

\break 





\section{Introduction}

\citeauthor{harsanyi1955}'s  (\citeyear{harsanyi1955}) aggregation theorem relates a utility function $u_{0}$, intended to represent the collective judgment of a group of $N$ individuals, to the utilities  $u_{i}$ of these individuals. In a setting with von Neumann-Morgenstern (vN-M) utilities over alternatives $x$, Harsanyi's theorem gives conditions that characterize the equality between $u_{0}$ and some utilitarian aggregation function
	\begin{align}\label{equation: formula w}
		w(x) = \sum_{i=1}^{N}a_{i}u_{i}(x)+b,
	\end{align}
by showing that $u_{0}$ is a utilitarian aggregation of the individual utilities $u_{i}$ when they satisfy a suitable Pareto unanimity principle.

While the standard Pareto principles characterize precise forms of utilitarian aggregation, economic models of behavior are only approximations of reality, and we should not expect their assumptions to hold exactly. In fact, many theoretical results in individual and collective decision-making rely on exact conditions that are false in many significant situations. This raises an important question about the extent to which the conclusions of these results can be preserved when their underlying assumptions are modified or weakened. For example, a decision maker (DM) seeking to provide a social assessment of competing alternatives may not enforce Pareto unanimity in every comparison, instead accepting alternatives that are merely near-optimal. Hence, when the precise conditions required by the Pareto unanimity principles are not met -- and so exact utilitarian aggregation does not obtain -- we seek conditions under which the strong conclusion of linear aggregation still holds, at least approximately.

This paper shows that the conclusions of Harsanyi’s aggregation theorem hold approximately when its assumptions are nearly satisfied. Our main goal is to define and characterize the notion of approximate utilitarian aggregation within an expected utility framework. We study how $u_{0}$, which we refer to here as the DM's utility rather than a social or group utility function, relates to individual utilities $u_{i}$ under weaker variants of the standard Pareto conditions. These variants relax the standard Pareto constraints, requiring only that the DM's choices be near-optimal rather than fully consistent with unanimous preferences. Allowing this form of satisficing behavior introduces flexibility into the aggregation procedure and permits us to quantify departures from utilitarianism. We show that the extent to which the Pareto principles are violated bounds the distance from $u_{0}$ to the set of utilitarian aggregators. Moreover, in many important settings, we obtain an exact quantitative relationship between this distance and the degree of violation of the Pareto principles. 

More specifically, under our near Pareto unanimity conditions, we prove that, rather than $u_{0}$ being identical to a utilitarian aggregator $w$ as in (\ref{equation: formula w}), a residual term $r(x)$ arises, so that 
\begin{align}\label{equation: u0 = w + r}
	u_{0}(x) =  w(x)+ r(x)\quad\text{for all }x.
\end{align} 
The magnitude of $r(x)$ in (\ref{equation: u0 = w + r}) is at most half of the positive parameter $\epsilon$ that bounds the extent to which the standard Pareto unanimity conditions are violated. Likewise, we show that our weaker Pareto principles are equivalent to a representation of the form
\begin{align}\label{equation: u0 = lin average + e}
	u_{0}(x)= \sum_{i=1}^{N}a_{i}u_{i}(x)+e(x)\quad\text{for all }x,
\end{align}
with $e$ denoting a function whose oscillation does not exceed that same $\epsilon$. This additional structure on the size of the residual term $r$ in (\ref{equation: u0 = w + r}) and, equivalently, on the oscillation of $e$ in (\ref{equation: u0 = lin average + e}) yields a precise upper bound on the distance of the DM's utility function to the set of utilitarian aggregators $w$. When the alternatives $x$ belong to certain domains that are common in economic applications, and the utility functions are sufficiently well-behaved, we exactly determine the distance of a non-utilitarian $u_{0}$ to the class of utilitarian aggregators $w$ by means of the violations of the Pareto conditions.

To illustrate, our version of the Semistrong Pareto principle requires that 
\begin{align}\label{equation: Semistrong intro}
	u_{0}(x)\geq u_{0}(y)-\epsilon \quad \text{ when }\quad u_{i}(x)\geq u_{i}(y) \text{ for all }i=1,\dots,N.
\end{align}
For interpretation, if alternative $x$ is an optimal choice from $\{x,y\}$ for each individual $i$, then the DM considers $x$ satisficing, meaning that $x$ is $\epsilon$-optimal over the set  $\{x,y\}$ in the sense of \cite{radner1980}. Note that in (\ref{equation: Semistrong intro}) we can have a preference for $y$ over $x$ for the DM, but this violation of the standard Pareto unanimity principle is bounded, or constrained, by the parameter $\epsilon$ in our cardinal setting. Hence, we allow some flexibility in the DM's preference for $x$ over $y$. Rather than requiring a preference for alternative $x$, the condition on the DM's preferences merely requires that $x$ should not be significantly inferior to $y$. More important, our results establish the equivalence between $\epsilon$-Pareto conditions as in (\ref{equation: Semistrong intro}) and representations of $u_{0}$ as an approximate utilitarian aggregator like in (\ref{equation: u0 = w + r}) with $\abs{r(x)}\leq \frac{\epsilon}{2}$ for every alternative $x$, and in (\ref{equation: u0 = lin average + e}) with $e(x)-e(y)\leq \epsilon$ for every pair of alternatives $x$ and $y$. 

The characterization of the $\epsilon$-Pareto axiom just described resembles a quantitative version of Harsanyi's theorem on utilitarian aggregation and, in a sense, demonstrates its stability. Specifically, small violations of the Pareto principle yield an aggregation rule that approaches linear aggregation. For settings with vN-M utilities that are common in many applications, we also establish in this paper a duality formula that connects both the norm of the residual term $r$ in (\ref{equation: u0 = w + r}) and the oscillation of the function $e$ appearing in (\ref{equation: u0 = lin average + e}) to violations of the standard Semistrong Pareto principle. In particular, the norm of the function $r$ and the oscillation of $e$ vanish precisely when the right-hand sides of (\ref{equation: u0 = w + r}) and (\ref{equation: u0 = lin average + e}) define a utilitarian aggregation function. At the same time, our measure of the violations of Semistrong Pareto equals zero in precisely this case.

When approximate utilitarian aggregation is viewed as a quantitative version of Harsanyi's aggregation theorem, this paper contributes to a significant literature that examines the extent to which a given object deviates from satisfying certain ideal assumptions and, consequently, from taking a specific form. One early example, in the context of the existence of macro production functions, is \cite{fisher1969}. He investigates the regularity conditions a function must satisfy to be close to a mapping in which two or more arguments are grouped into a single aggregate.\footnote{Fisher's question is addressed in further detail in \cite{mak1988}, who gives a sharper quantitative result linking the degree of near separability of a function to its deviation from a truly separable mapping.} More recently, results of a similar kind have appeared in \citeauthor{mossel2012}'s (\citeyear{mossel2012}) quantitative version of Arrow’s impossibility theorem, as well as in the works of \cite{hellman2013} and \cite{hellmanpinter2022}, who explore how deviations from the common priors assumption among a group of agents can be quantified through the presence of positive gains from betting. In the finance literature, \cite{acciaioetal2022} establish a connection between the magnitude of normalized arbitrage opportunities and the extent to which a pricing functional is approximately a martingale measure. Similarly, using de Finetti’s definition of arbitrage opportunities, \cite{nascimento2024} derives a related result for the aggregation of probability measures and the representation of nearly rational stochastic choice functions.\footnote{For a recent account of de Finetti's theorem, see \cite{nielsen2019,nielsen2021}.}

This paper is organized as follows. Section \ref{section: framework} introduces our setting with vN-M utilities and presents equivalent definitions of approximate utilitarian aggregation. Section \ref{section: aggregation finite} establishes the main theorems, both with and without sign restrictions on Pareto weights, and explores two applications to aggregating subjective probabilities and tastes. While Section \ref{section: conclude} concludes with remarks on our approach and open questions, for ease of exposition all proofs are deferred to Appendix \ref{appendix}.


\section{Framework}\label{section: framework}

The set $X$ is the set of alternatives on which preferences are defined. We assume that $X$ is a nonempty subset of a real vector space. We also take as primitive a collection of complete and transitive binary relations (preference orderings) on $X$. From this collection, we distinguish a preference ordering $\pref_{0} \,\subseteq X\times X$ for the DM and a finite set $\{\pref_{1},\dots,\pref_{N}\}$ of binary relations on $X$. Here, $\pref_{i}$ denotes the preferences of individual $i$. The mappings $u_{0}\colon X \to\Re$ and $u_{i}\colon X\to\Re$ are fixed throughout as representations of $\pref_{0}$ and $\pref_{i}$, respectively, for $i=1,\dots, N$. For notation, we also define $\U$ as the set of all such $u_{i}$, namely, 
\begin{align}\label{equation: set U finite}
\U = \{u_{1},\dots,u_{N}\}.	
\end{align}
We sometimes refer to a utility function in $\U$ as $u$, thus omitting its subscript. By convention, we use the subscript $j$ to refer either to an individual $i = 1, \dots, N$ or to the DM (with $j = 0$).

Our main results require additional structure on the set of alternatives and on the utility functions. To this end, we define a particular instance of the setting in which $X$ can be interpreted as a set of lotteries and the utility functions have the relevant structure shared by expected utility functionals.

\begin{definition}
	The triple $(X,u_{0},\U)$ is called a setting with vN-M utilities when the set $X$ is also convex, and the utility functions $u_{0}$ and $u_{i}\in \U$, $i=1,\dots,N$, are affine mappings, with the meaning that they have the mixture-preserving property $u_{j}(\lambda x+ (1-\lambda)y)=\lambda u_{j}(x) + (1-\lambda) u_{j}(y)$ for all $x,y\in X,$ and $0\leq\lambda \leq 1$.
\end{definition} 

We now present essential examples of settings with vN-M utilities.

 \begin{example}[Expected Utility 1]\label{example EU 1}
 	The set $O$ represents a basic set of deterministic outcomes or prizes. The elements of $X$ are the lotteries on $O$ with a finite number of outcomes, namely, the functions $p\colon O\to[0,1]$ such that the set $\{o\in O:p(o)\neq 0\}$ is finite, with $\sum_{o\in O}p(o)=1$. Therefore, the set $X$ of simple lotteries is a convex subset of $\Re^{O}$. Utility functions have the form $p\mapsto\sum_{o\in O}p(o)v(o)$ for some $v\colon O\to\Re$. \myqed
 \end{example}
 
\begin{example}[Expected Utility 2]\label{example EU 2}
The set $O$ is the same as in Example \ref{example EU 1}, and  $\Sigma_{O}$ is a $\sigma$-algebra of its subsets. The set  $X$ consists of all probability measures on $O$, i.e., countably additive functions $P\colon \Sigma_{O}\to [0,1]$ with $P(O)=1$. Then $X$ is a convex subset of the vector space of all signed measures of bounded variation on the measure space $(O,\Sigma_{O})$. The relevant utility functions are expected utility functionals, mapping $P\in X$ to  $\int_{O}v(o)dP(o)$ for some bounded and $\Sigma_{O}$-measurable function $v\colon O\to\Re$. \myqed
\end{example}

\begin{example}[Subjective Expected Utility 1]\label{example SEU 1}
Given a set $S$ of states, an algebra $\A$ of its subsets, and a convex subset $C$ of a real vector space, the set $X$ of alternatives consists of all acts $f\colon S\to C$ with finite range and for which $f^{-1}(c)\in\A$ for all $c\in C$. Here $C$ is interpreted as the set of consequences, and utility functions map each act $f$ to $\int_{S}v(f(s))d\mu(s)$, where the function $v\colon C\to \Re$  is affine and  $\mu\colon \A \to [0,1]$ is a subjective finitely additive probability measure.\myqed
\end{example}

\begin{example}[Subjective Expected Utility 2]\label{example SEU 2}
Let $S$ denote a set of states as in Example \ref{example SEU 1}, $\Sigma_{S}$ a  $\sigma$-algebra on $S$, and $C$ a set of consequences. For  some $c^{\ast},c_{\ast}\in C$ we have the strict preference $c^{\ast}\succ_{j} c_{\ast}$ for all $j=0,1,\dots,N$. The set of alternatives consists of all acts $f\colon S\to C$ with finite range and for which $f^{-1}(c)\in  \Sigma_{S}$ for all $c\in C$. For each $j$, there exists a function  $v_{j}\colon C\to\Re$ and a nonatomic probability measure $P_{j}\colon\Sigma_{S}\to [0,1]$ such that $u_{j}(f) = \sum_{c\in C}P_{j}(f^{-1}(c))v_{j}(c)$.\footnote{Recall that a probability measure $P$ on $\Sigma_{S}$ is non-atomic if for all $E_{1}\in \Sigma_{S}$ with $P(E_{1})>0$ there exists $E_{2}\subseteq E_{1}$ with $E_{2}\in \Sigma_{S}$ and $0<P(E_{2})<P(E_{1})$.} By Lyapunov's theorem, the convex set $\mathcal L(C)$ of all simple lotteries $p$ on $C$ is identified with a subset of acts, and for each $j$,  $u_{j}(p) = \sum_{c\in C}p(c)v_{j}(c) $ where $p(c) = P_{j}(f^{-1}(c))$ represents the agreed-upon probability among the DM and individuals. The restrictions of $u_{0}$ and $u\in \U$ to $\mathcal L(C)$ define a setting with vN-M utilities. Lyapunov’s theorem also ensures that the set of probability profiles $\bm P(E) = (P_{0}(E),P_{1}(E),\dots,P_{N}(E))$ is a convex subset of $\Re^{N+1}$. The restriction of utility functions to binary acts $f $, where $f(s) = c^{\ast}$ for $s\in E$ and $f(s) = c_{\ast}$ otherwise,  induces preferences on $\bm P(E)$ represented by $\bm P(E)\mapsto (v_{j}(c^{\ast})-v_{j}(c_{\ast}))P_{j}(E) + v_{j}(c_{\ast})$. This procedure also induces a setting with vN-M utilities. \myqed 
\end{example}

In the case of a finite set of prizes, Example \ref{example EU 1} corresponds to the standard expected utility framework of \cite{vnm1947}.\footnote{See, e.g., the works of \cite{borgers2017} and \cite{karni2024} on relative utilitarianism for a recent use of this setting.} Example \ref{example EU 2} specializes to a framework with a richer structure on the set of prizes and corresponds to the setting of \cite{border1985} in the context of (exact) utilitarian aggregation. Meanwhile, Example \ref{example SEU 1} covers the canonical Anscombe-Aumann setting, as considered in \cite{mongin1995,mongin1998}. Lastly, the two vN-M utility settings identified in Example \ref{example SEU 2} play a crucial role in the aggregation results of \cite{gilboa2004} and \cite{alon2016,alon2024} within \citeauthor{savage1954}’s (\citeyear{savage1954}) framework.

It is also convenient to distinguish the cases where the domain $X$ has the topological property of compactness, as is typical in common and important economic applications, and where the utility functions involved are continuous. While this is made precise in the next definition, we refer the reader to \cite{aliprantis2006} for the technical background.  

\begin{definition}
	The setting with vN-M utilities $(X,u_{0},\U)$ is called a setting with continuous vN-M utilities and a compact domain if the set $X$ is also a compact subset of a Hausdorff topological vector space, and the utility functions $u_{0}$ and $u_{i}\in \U$, $i=1,\dots,N$, are also continuous.
\end{definition}

To see this concretely, Example \ref{example EU 1} constitutes a setting with continuous vN-M utilities and a compact domain when the set $O$ is finite. Example \ref{example EU 2} also has this property when $O$ is a compact metric space, and the set of (Borel) probability measures on $O$ is endowed with the weak$^{\ast}$ topology. In this setting, continuous utility functions are obtained by integrating a continuous utility function over final outcomes in $O$ with respect to a probability measure. Example \ref{example SEU 1} also yields a setting with continuous vN-M utilities and a compact domain when the set of states is finite, and consequences are lotteries over finitely many prizes.

For future reference (see Section \ref{section SEU examples}), in the context of Example \ref{example SEU 2}, it is useful to define the set $\Sigma_{u}\subseteq \Sigma_{S}$ as the collection of all events with a unanimously agreed-upon probability:
\begin{align}\label{equation: sigmau defined}
	 \Sigma_{u} = \{E\in\Sigma_{S}:P_{0}(E)=P_{i}(E)\text{ for all }i=1,\dots,N\}.
\end{align}
 Also, for a signed measure of bounded variation $R$  on $\Sigma_{S}$, we define its total variation norm with the formula $\norm{R}_{1} = R^{+}(S)+R^{-}(S)$, where $R^{+}$ and $R^{-}$ are nonnegative measures on $\Sigma_{S}$ such that $R = R^{+}-R^{-}$. The existence of $R^{+}$ and $R^{-}$ is assured by the Jordan decomposition, and we have $R^{+}(S)=\sup_{E\in\Sigma_{S}}R(E)$ and  $R^{-}(S)=-\inf_{E\in\Sigma_{S}}R(E)$; see, e.g., \citet[pp.~420-421]{billingsley1995}.

We also recall that the supremum norm of a real-valued and bounded function $h$ with domain $D$ is
\[\norm{h}_{\infty} =\sup_{d\in D}\abs{h(d)}.
\]
The oscillation of $h$ on its domain $D$ is the nonnegative real number
\[\omega_{h}(D) = \sup_{d\in D}h(d)-\inf_{d^{\prime}\in D}h(d^{\prime}).
\]
Equivalently, the oscillation of $h$ is the supremum of the distance between any two values in its image, namely,  $ \sup_{d,d^{\prime}\in D}\abs{h(d)-h(d^{\prime})}$. For notation, given a set $D$, and $E\subseteq D$, we define the indicator function $\indic_{E}\colon D\to\Re$ with $\indic_{E}(d)=1$ if $d\in E$, and $\indic_{E}(d)=0$ otherwise. Also, we denote by $\Re_{+}$ the set of all nonnegative real numbers.

Our characterization results in the next section identify all pairs $(u_{0},\U)$ in a vN-M setting where $u_{0}$ is approximately a utilitarian aggregator of the utilities in $\U$. This requires the definition below. 

\begin{definition}\label{definition: utilitarian aggregators}
	A utilitarian aggregator of the functions in $\U$ is a function $w\colon X\to\Re$ such that there exist $a_{1},\dots,a_{N}$ and $b$ in $\Re$ with 
	\begin{align}\label{equation: define utilitarian aggregator w}
		w(x)=\sum_{i=1}^{N}a_{i}u_{i}(x)+b\quad\text{for all }x\in X.
	\end{align}
The set of all utilitarian aggregators of the elements of $\U$ is denoted by $\W$. We also define $\W_{+}$ as the subset of $\W$ of those utilitarian aggregators with $a_{i}\geq 0$ for all $i=1,\dots,N$, and $\W_{++}$ as the further subset of those $w$ with all $a_{i}>0$. 
\end{definition} 

According to Definition \ref{definition: utilitarian aggregators}, a utilitarian aggregator is a weighted sum $\sum_{i=1}^{N}a_{i}u_{i}+b\indic_{X}$ of the individual utilities and the indicator function of $X$. The constant function $b\indic_{X}$ in the expression is a function with no oscillation. As anticipated above in equation (\ref{equation: u0 = w + r}), our approximately utilitarian results allow for an additional residual term $r$, which must be distinguished from that constant. In light of this, the following proposition presents two equivalent forms of what it means for $u_{0}$ to be approximately a utilitarian aggregator of the elements of $\U$. Its proof relies on a straightforward geometric idea: any function with bounded oscillation can be adjusted to be centered at zero, reducing its supremum norm to half of its oscillation.

\begin{proposition}\label{proposition: equivalent forms of approximation}
	Suppose that $\epsilon\geq 0$. The following statements are equivalent.
	\begin{enumerate}
		\item[(i)] There exist real numbers $a_{1},\dots,a_{N}$ and a function $e\colon X\to \Re$ with $\omega_{e}(X)\leq \epsilon$, such that $u_{0}(x) = \sum_{i=1}^{N}a_{i}u_{i}(x) +e(x)$ for all $x\in X$. 
		\item[(ii)] The function $u_{0}$ is $\frac{\epsilon}{2}$-close to a utilitarian aggregator, that is, for some $w\in\W$, we have that $\norm{u_{0} - w}_{\infty}\leq\frac{\epsilon}{2}$.
	\end{enumerate}
	Moreover, the equivalence also holds when we assume that $a_{i}\geq 0$ (respectively, $a_{i}>0$) for all $i=1,\dots,N$, and at the same time $\W$ is replaced by $\W_{+}$ (respectively, $\W_{++}$).
\end{proposition}

Proposition \ref{proposition: equivalent forms of approximation} says that for a fixed $\epsilon\geq 0$, we have two equivalent forms of viewing the utility function $u_{0}$ as approximately utilitarian. The first views $u_{0}$ as the linear average of the functions $u_{i}$ plus a function $e$ whose oscillation is bounded by $\epsilon$. Note that in the case of $\epsilon=0$ we are in a fully utilitarian setting where the function $e$ is constant, so $u_{0}$ coincides with an element of $\W$. The second states that, as mentioned in the Introduction and repeated below for convenience, $u_{0}$ is at a distance from the set $\W$ bounded by $\frac{\epsilon}{2}$, meaning that
\begin{align}\label{equation: decompose u0}
	u_{0}(x) = w(x)+r(x)
\end{align}
for some function $r$ where $\norm{r}_{\infty}\leq \frac{\epsilon}{2}$. Here we note that in a setting with vN-M utilities, the expression in (\ref{equation: decompose u0}) presents a decomposition of the utility function $u_{0}$ into an affine function $w$ that is a utilitarian aggregator, and another affine function $r$ which is not a primitive in the aggregation problem. In our cardinal setting, the idea is that the parameter $\epsilon$ controls the importance of the residual term $r$ on the right-hand side of (\ref{equation: decompose u0}).

The equivalence derived in Proposition \ref{proposition: equivalent forms of approximation} between the two properties of the utility functions motivates the following definition.

\begin{definition}\label{definition: utilitarian aggregation with norm}
	Let $\epsilon\geq 0$. The utility function $u_{0}$ is called approximately utilitarian given $(\epsilon, \W)$ (respectively, $(\epsilon, \W_{+})$ or $(\epsilon, \W_{++})$) if there exists a utilitarian aggregator $w\in \W$ (respectively, $w\in \W_{+}$ or $w\in \W_{++}$) such that $\norm{u_{0}-w}_{\infty}\leq \frac{\epsilon}{2}$.
\end{definition}

\begin{remark}\label{remark: relative size epsilon}
	In the standard setting of utilitarian aggregation with $\epsilon=0$, at least one of the Pareto weights is nonzero as long as the asymmetric part of the preference ordering $\pref_{0}$ is nonempty. In a setting with approximately utilitarian aggregation, a similar conclusion holds with $\W$ and $\W_{+}$ provided that  $\epsilon< \omega_{u_{0}}(X)$. This property of the utilitarian aggregator $w$ that is $\frac{\epsilon}{2}$-close to $u_{0}$ follows from the observation that $\omega_{u_{0}}(X)=\omega_{e}(X)$ if $a_{i} = 0$ for all $i=1,\dots,N$. \myqed
\end{remark}

\begin{remark}
As in the case of utilitarian aggregation with the standard Pareto axioms, the approximate versions of the classic results, such as those in Definition \ref{definition: utilitarian aggregation with norm} or in their equivalent forms given in Proposition \ref{proposition: equivalent forms of approximation}, depend only on the cardinal representations $u\in\U$ of the preferences of the individuals. More concretely, suppose there exist constants $\alpha_{i}>0$ and $\beta_{i}\in\Re$ such that $\hat u_{i}=\alpha_{i}u_{i}+\beta_{i}$ for all $i=1,\dots,N$. If $u_{0}$ is approximately utilitarian in any of the senses described in Definition \ref{definition: utilitarian aggregation with norm}, the same conclusion holds when the functions $u_{i}$ are replaced with the cardinally equivalent utility functions $\hat  u_{i}$.\footnote{Specifically, whenever $e = u_{0}-\sum_{i=1}^{N}a_{i}u_{i}$ has  $\omega_{e}(X)\leq \epsilon$, the oscillation of the function  $\hat  e = u_{0}-\sum_{i=1}^{N}\hat  a_{i}\hat u_{i}$ is not greater than $\epsilon$ when $\hat  a_{i} = \frac{a_{i}}{\alpha_{i}}$, since $\omega_{\hat  e}(X)=\omega_{e}(X)$.} Therefore, common normalizations of the utilities in $\U$, such as those used in the relative utilitarianism literature (e.g., \cite{borgers2017, karni2024}) or those that fix a prescribed value for a given pair of alternatives for each individual, yield the same class of $\epsilon$-approximate utilitarian representations. This invariance, however, fails if the cardinal representation of the DM's preferences changes scale, that is, if $u_{0}$ is replaced with $\alpha_{0}u_{0}+\beta_{0}$ for $\alpha_{0}>0$ and $\alpha_{0}\neq 1$.  \myqed
\end{remark}


\section{Approximate utilitarian aggregation}\label{section: aggregation finite}



\subsection{Approximate aggregation with nonnegative weights}


We begin by considering a form of approximate utilitarian aggregation based on a modified Semistrong Pareto condition.\footnote{For terminology regarding the Pareto axioms, see \cite{weymark1991,weymark1994}.} In its original version, this condition states that if $x$ and $y$ are alternatives in $X$ such that $x$ is an optimal choice from $\{x,y\}$ for every individual, then $x$ is also an optimal choice for the DM from that two-element set. In contrast, we relax the constraints imposed by the Semistrong Pareto condition, requiring only that the DM's choices be $\epsilon$-optimal when the individual utilities provide a unanimous ranking (see Definition \ref{definition: epsilon Semistrong} below).

The notion of $\epsilon$-optimality requires comparing utility differences to a given threshold. To operationalize this comparison using only preferences in our setting with mixture-preserving utilities -- and to control the degree of violation of the standard Pareto axioms with a single parameter $ \epsilon$ -- we first fix a pair of strictly ranked alternatives, $\xh\succ_{0}\xl$. We then use the difference $u_{0}(\xh)-u_{0}(\xl)$ as a unit of measurement, with $\epsilon$ defining the threshold as a fraction of this unit. For $\epsilon \geq 0$, the near-optimality of $x$ relative to $y$ may be taken to mean that
\begin{align}\label{equation: nearly preferable}
	\frac{1}{1+\epsilon} x +  \frac{\epsilon}{1+\epsilon} \xh \pref_{0}  \frac{1}{1+\epsilon} y +  \frac{\epsilon}{1+\epsilon}\xl.
\end{align}
In the special case where $\epsilon=0$, near-optimality corresponds to the weak preference $x\pref_{0}y$, as in the Semistrong Pareto axiom. Otherwise, with $\epsilon>0$ and small, we may have $y\succ_{0}x$ but the  convex combination of $x$ and  $\xh$ with weights $\frac{1}{1+\epsilon} $ and $\frac{\epsilon}{1+\epsilon} $, respectively, breaks the strict preference when $y $ is combined with alternative $\xl$, using the same respective weights as $x$ and $\xh$. This is what the weak preference in (\ref{equation: nearly preferable}) expresses.\footnote{Alternatively, as in the discussion following Definition \ref{definition: sequential epsilon strong} below, we can dispense with $\epsilon$ in the formulation of $\epsilon$-optimality using preferences by assuming that $\xh$ and $\xl$ are, respectively, the best and worst alternatives for the DM in $X$. In this case, consider alternatives $x^{\ast}$ and $x_{\ast}$ such that $\xh \pref_{0} x^{\ast}\succ_{0} x_{\ast}\pref_{0}\xl$ and $u_{0}(x^{\ast})-u_{0}(x_{\ast}) = \epsilon [u_{0}(\xh)-u_{0}(\xl)]$. Here the relevant $\epsilon$ lies in the interval $ [0,1]$. Then the near preference for $x$ over $y$ in (\ref{equation: nearly preferable}) can be replaced by $\frac{1}{2} x +  \frac{1}{2} x^{\ast} \pref_{0}  \frac{1}{2} y +  \frac{1}{2}x_{\ast}$.}

Since $u_{0}$ is an expected utility function in our setting with vN-M utilities, the comparison in (\ref{equation: nearly preferable}) requires that the utility difference $u_{0}(y) - u_{0}(x)$ be at most $[u_{0}(\xh)-u_{0}(\xl)]\epsilon$. The quantity $u_{0}(\xh)-u_{0}(\xl)$ thus serves as a unit of measurement for expressing near-optimal choices, while the parameter $\epsilon$ controls the degree of such approximate optimality. Moreover, if we further assume that utility differences reflect the strength of preference between pairs of alternatives, then $\epsilon$-optimality incorporates the idea that choosing $x$ from $\{x,y\}$ is admissible even when $y$ is strictly preferred to $x$, provided the strength of this preference does not exceed $\epsilon$ times the ``degree of preference'' for $\xh$ over $\xl$.\footnote{This interpretation is controversial and has been the subject of considerable debate. We refer the reader to \citet[pp.\ 80-86]{fishburn1970} for a thoughtful early discussion of the notion of strength of preference.}
 
For practical purposes, we assume that the choices of $\xh$ and $\xl$ are implicit in each context and formulate our Pareto condition solely in terms of utilities, referring only to $\epsilon$, which represents the relevant threshold. This is described below.

\begin{definition}[$\epsilon$-Semistrong Pareto]\label{definition: epsilon Semistrong}
	The pair $(u_{0},\U)$ satisfies the $\epsilon$-Semistrong Pareto condition when for all $x,y\in X$: if $u_{i}(x) \geq  u_{i}(y)$ for all $i=1,\dots,N$, then $u_{0}(x)\geq u_{0}(y) -\epsilon$.
\end{definition}

To illustrate, when the utility $u_{0}$ is normalized so that $u_{0}(\xh)=1$ and $u_{0}(\xl)=0$ -- a common normalization -- the conclusion of $\epsilon$-Semistrong Pareto, given a unanimous ranking of $x$ and $y$, is equivalent to (\ref{equation: nearly preferable}). Otherwise, caution is required when interpreting the conclusion of $\epsilon$-Semistrong Pareto with preferences. As noted above, in settings with vN-M utilities the notion of near-optimality formulated in (\ref{equation: nearly preferable}) becomes equivalent to $u_{0}(x)\geq u_{0}(y)-[u_{0}(\xh)-u_{0}(\xl)]\epsilon$.\footnote{A related interpretation of the $\epsilon$-Pareto axiom can be given in a setting with preferences over lotteries defined on a compact interval of monetary outcomes. In this context, $\epsilon$ can be expressed in terms of outcomes, and the near-optimality condition can be reformulated accordingly, using comparisons of certainty equivalents.} 

At the same time, as noted in Remark \ref{remark: relative size epsilon} and \textit{referring only to utilities} as in Definition \ref{definition: epsilon Semistrong}, when $\xh$ and $\xl$ represent the best and worst alternatives in $X$ for the DM, the $\epsilon$-Semistrong Pareto axiom holds trivially if $\epsilon\geq u_{0}(\xh)-u_{0}(\xl)$. To avoid trivialities in such cases, we must have $0\leq \epsilon<u_{0}(\xh)-u_{0}(\xl)$. More generally, without such an assumption about the best and worst alternatives, we require $\epsilon<\omega_{u_{0}}(X)$ for the $\epsilon$-Semistrong Pareto condition to impose a meaningful restriction.

To characterize our weakening of the Semistrong Pareto condition in a cardinal setting, we first consider the case of a setting with continuous vN-M utilities and a compact domain. The additional structure on the set of alternatives and the utility functions allows us to establish a quantitative utilitarian aggregation theorem that relates the DM's cardinal utility representation to a linear average of the individual utilities. Such a theorem provides more information about how the DM's preferences relate to those of the individuals than our goal of finding an approximately utilitarian aggregation rule with $\epsilon$-Semistrong Pareto. 

To this end, we first define the set of pairs of alternatives that correspond to potential violations of Semistrong Pareto. This is the set
\begin{align}\label{equation: V defined}
	\V = \left\{(x,y)\in X\times X: u_{0}(y)\geq u_{0}(x) \text{ and }u_{i}(x)\geq u_{i}(y)\text{ for all }i=1,\dots,N\right\}.
\end{align}
Under Semistrong Pareto, the set $\V$ is a subset of the symmetric part of the DM's preference relation and contains the diagonal in $X\times X$. In our weaker version of such an axiom, the set $\V$ may have a richer structure. More generally, the set $\V$ is in duality with the difference between $u_{0}$ and the linear averages of the $u_{i}$'s. These  differences are induced by vectors of weights $\bm a = (a_{1},\dots,a_{N})$ with $a_{i}\geq 0$, and have the form $u_{0}-\sum_{i=1}^{N}a_{i}u_{i}$. This is shown in the theorem below.

\begin{theorem}\label{theorem: duality V}
	Suppose that  $(X,u_{0},\U)$ is a setting with continuous vN-M utilities and a compact domain, and that $\V$ is the subset of $X\times X$ defined in (\ref{equation: V defined}). Then, there exists some $\bm a = (a_{1},\dots,a_{N})\in \Re^{N}_{+}$ such that the mapping $e\colon X\to\Re$, defined by 
	\[
	u_{0}(x) = \sum_{i=1}^{N}a_{i}u_{i}(x)+e(x),
	\]
	satisfies
	\begin{align}\label{equation: duality V}
		\omega_{e}(X) = \max\{u_{0}(y)-u_{0}(x):(x,y)\in\V \}.
	\end{align}

\end{theorem}

Theorem \ref{theorem: duality V} is essentially a utilitarian aggregation result without the Pareto axioms. Its proof shows that a special function, defined as the difference (in their arguments) of differences of the form $w(x) - u_{0}(x)$, with $w$ ranging over $\W_{+}$, is sufficiently well-behaved so that a suitable version of the minimax theorem can be applied. This application enables us to eventually find the Pareto weights that make the duality in (\ref{equation: duality V}) hold by showing that the minimum over $w\in\W_{+}$ of the maximum over $x$ and $y$ of
\begin{align}\label{equation: differences of differences}
	w(x) - u_{0}(x) + u_{0}(y) - w(y)
\end{align}
is attained. This part strongly requires that the set of utilities $\U$ be finite. With respect to the utility functions themselves, the argument also reveals that the minimum distance between $u_{0}$ and the set $\W_{+}$ is attained with the same particular choice of weights $a_{i}\geq 0$. The equivalence between the two tasks follows from the observation that the constant $b$ in the utilitarian aggregator, as in equation (\ref{equation: define utilitarian aggregator w}), plays no part in the definition of the differences in (\ref{equation: differences of differences}). 

 More important, Theorem \ref{theorem: duality V} establishes a connection between the utility differences $u_{0}(y)-u_{0}(x)\geq 0$ for $(x,y)\in \V$ -- which can be interpreted as measuring violations of the Semistrong Pareto axiom -- and the oscillation of the function $e$ in the approximate utilitarian aggregation described in part (i) of Proposition \ref{proposition: equivalent forms of approximation}. It requires no form (weaker or otherwise) of Pareto unanimity, but rather quantifies the distance between the functions in the set $\W_{+} $ of aggregators and the DM's utility by measuring the strength of violations of the Semistrong Pareto condition. 
 
 When all utility differences in the set $\V$ are zero, meaning the Semistrong Pareto axiom is satisfied, the function $e$ remains constant because its oscillation is zero. Here we arrive at Harsanyi's conclusion that $u_{0}\in W_{+}$. Otherwise, due in particular to the existence of a utilitarian aggregator in $\W_{+}$ at a minimum distance from $u_{0}$, as mentioned in the paragraph following Theorem \ref{theorem: duality V}, the idea of the proof of Proposition \ref{proposition: equivalent forms of approximation}, specifically Lemma \ref{lemma: auxiliary geometric} in Appendix \ref{appendix}, also implies that a utilitarian aggregator $\sum_{i=1}^{N}a_{i}u_{i}+b\indic_{X}\in \W_{+}$ can be chosen to satisfy 
 \begin{align}\label{equation: duality distance fixed w oscillation}
 	 \norm{r}_{\infty}= \frac{\omega_{e}(X)}{2},
 \end{align}
 where $r =u_{0}-\sum_{i=1}^{N}a_{i}u_{i}-b\indic_{X} $ is the residual term referred to in equation (\ref{equation: u0 = w + r}) in the Introduction. In fact, we obtain the duality formula
\begin{align}\label{equation: duality using distances}
	\min_{w\in W_{+}}\norm{u_{0}-w}_{\infty} =  \frac{\max\{u_{0}(y)-u_{0}(x):(x,y)\in\V \}}{2},
\end{align}
that is, the distance between  $u_{0}$ and the set of utilitarian aggregators $\W_{+}$ is half of the maximum cardinal ``intensity'' of the violations of the Semistrong Pareto axiom, as captured by the right-hand side of equation (\ref{equation: duality V}).\footnote{Here we refer the reader to Lemma \ref{lemma: minimax duality} in Appendix \ref{appendix}, where we show that the minimum, over nonnegative weight vectors $\bm a$, of the maximum value (in $x$ and $y$) of the expression given in (\ref{equation: differences of differences}) equals the oscillation appearing in equation (\ref{equation: duality V}). Hence, in light of the discussion following the expression in (\ref{equation: differences of differences}) about the constant $b$, we cannot have a utilitarian aggregator in $\W_{+}$ whose distance to $u_{0}$ is less than the value specified in equation (\ref{equation: duality distance fixed w oscillation}). This establishes the duality formula in (\ref{equation: duality using distances}).}  

We now employ the duality established in Theorem \ref{theorem: duality V} to show our aggregation result with $\epsilon$-Semistrong Pareto in the more restricted setting of Theorem \ref{theorem: duality V}.   

\begin{theorem}\label{theorem: aggregation finite semistrong compact}
		Let $\epsilon\geq 0$, and suppose that  $(X,u_{0},\U)$ is a setting with continuous vN-M utilities and a compact domain. The following statements are equivalent.
	\begin{enumerate}
		\item[(i)] The pair  $(u_{0},\U)$ satisfies $\epsilon$-Semistrong Pareto.
		
		\item[(ii)] The utility function $u_{0}$ is approximately utilitarian given $(\epsilon,\W_{+})$.
	\end{enumerate}	
\end{theorem}

Theorem \ref{theorem: aggregation finite semistrong compact} is a straightforward application of Theorem \ref{theorem: duality V}. Its proof therefore indirectly relies on the minimax theorem, in the form established in the proof of  Theorem \ref{theorem: duality V}. The key assumption that enables its application is the additional topological structure, which is crucial to the proof of Theorem \ref{theorem: duality V}. The contribution of Theorem \ref{theorem: aggregation finite semistrong compact} lies in providing an explicit bound on the violations of the Semistrong Pareto axiom, as reflected in the use of the parameter $\epsilon \geq 0$ in Definition \ref{definition: epsilon Semistrong}. As noted earlier, this bound captures the $\epsilon$-optimality of the DM in the presence of unanimity.

We can now apply Theorem \ref{theorem: aggregation finite semistrong compact} to obtain a more general version of the approximate utilitarian aggregation rules. Our next result eliminates topological assumptions, thereby focusing on the broader setting with vN-M utilities, and uses an extension argument to derive the relevant Pareto weights. 

\begin{theorem}\label{theorem: semistrong without compactness}
	Let $\epsilon\geq 0$, and suppose that  $(X,u_{0},\U)$ is a setting with vN-M utilities. The following statements are equivalent.
	\begin{enumerate}
		\item[(i)] The pair $(u_{0},\U)$ satisfies $\epsilon$-Semistrong Pareto.
		\item[(ii)] The utility function $u_{0}$ is approximately utilitarian given $(\epsilon,\W_{+})$.
	\end{enumerate}
\end{theorem}

The proof of Theorem \ref{theorem: semistrong without compactness} uses the result from Theorem \ref{theorem: aggregation finite semistrong compact} and applies an extension argument to characterize the approximate utilitarian aggregation rule over the entire set of alternatives. The key idea is to recognize that Theorem \ref{theorem: aggregation finite semistrong compact} applies to subsets of $X$ that are roughly expressed as lotteries over finitely many prizes, and to make a good choice of the finitely many points in $X$ to pin down the Pareto weights for each such set of finite lotteries using simple matrix algebra. The Pareto weights that make the statement in part (ii) of Theorem \ref{theorem: semistrong without compactness} hold are obtained using a limiting argument.

To interpret the result, the Introduction outlined a perspective based on the stability of Harsanyi’s conclusion regarding linear aggregation under bounded violations of the Pareto principles. Theorem \ref{theorem: semistrong without compactness} formalizes this idea: the parameter $\epsilon$ bounds not only the extent to which the standard Pareto axioms are violated but also the distance of $u_{0}$ from a utilitarian aggregator in $\W_{+}$.

We now turn to the Pareto principle as it relates to indifference. According to Harsanyi's Pareto indifference condition, the DM is indifferent whenever all individuals are. In our weaker version of Pareto indifference, the DM is nearly indifferent whenever individuals are indifferent.
  
\begin{definition}[$\epsilon$-Pareto Indifference]\label{definition: epsilon Pareto indifference}
	The pair $(u_{0},\U)$ satisfies $\epsilon$-Pareto Indifference when for all $x,y\in X$: if $u_{i}(x) = u_{i}(y)$ for all $i=1,\dots,N$, then $u_{0}(x)\geq u_{0}(y) -\epsilon$.
\end{definition}

Our Definition \ref{definition: epsilon Pareto indifference} has two immediate implications. First, it allows for the possibility that the DM's preferences are not fully aligned with those of the individuals in the group. Here, the special case $\epsilon=0$ corresponds to Harsanyi's Pareto Indifference axiom. Second, given the symmetry of the indifference relations in Definition \ref{definition: epsilon Pareto indifference}, we obtain an equivalent formulation of $\epsilon$-Pareto Indifference:
\begin{align*}
	\abs{u_{0}(x)-u_{0}(y)}\leq \epsilon\quad\text{ when }\quad u_{i}(x) = u_{i}(y)\text{ for all }i=1,\dots,N.
\end{align*}
This implies that while the DM need not be indifferent between alternatives $x$ and $y$, the utility values $u_{0}(x)$ and $u_{0}(y)$ must be close to each other.

As the following theorem shows, our weaker version of Pareto Indifference is equivalent to  $u_{0}$ being approximately utilitarian. It follows directly from Theorem \ref{theorem: semistrong without compactness} by adding, for each $u\in\U$, $-u$ to the set of individual utilities.

\begin{theorem}\label{theorem: aggregation finite indifference}
	Let $\epsilon\geq 0$, and suppose that  $(X,u_{0},\U)$ is a setting with vN-M utilities. The following statements are equivalent.

	\begin{enumerate}
		\item[(i)] The pair  $(u_{0},\U)$ satisfies $\epsilon$-Pareto Indifference.
		
		\item[(ii)] The utility function $u_{0}$ is approximately utilitarian given $(\epsilon,\W)$.
	\end{enumerate}
\end{theorem}

The example below illustrates our concept of approximately utilitarian representation in situations where the standard Pareto Indifference axiom does not hold. First, it identifies a case in which the approximate versions of the Pareto axioms introduced above fail whenever $\epsilon$ is smaller than the oscillation of the DM’s utility function. In this situation, only trivial forms of approximate utilitarian aggregation obtain, since $\epsilon \geq \omega_{u_{0}}(X)$ is required. Second, for a slight variation of the DM's utility that also leads to a failure of the standard Pareto Indifference axiom, it shows that an approximate form of aggregation is nevertheless still possible.

\begin{example}
	Suppose that $X= \{(x_{1},x_{2},x_{3})\in \Re^{3}: x_{1}+x_{2}+x_{3}=1,x_{1}\geq 0,x_{2}\geq 0,x_{3}\geq 0\}$,  $N=2$, and $\alpha\in [0,1)$ is a given parameter. For $x=(x_{1},x_{2},x_{3})$, utilities are $u_{0}(x) = x_{3}+\alpha x_{2}$, $u_{1}(x)= x_{1}$, and $u_{2}(x)=1-x_{1}$. We assume $0<\epsilon<\omega_{u_{0}}(X)=1$ to avoid trivialities, since an approximate utilitarian aggregation is always possible (as remarked in the discussion following Definition \ref{definition: epsilon Semistrong}) when $\epsilon\geq \omega_{u_{0}}(X)$. When $\alpha = 0$, the example corresponds to the one given in \citet[pp.\ 268-269]{weymark1991}, where the individuals have preferences with opposite directions, and the DM has preferences that are nontrivial and do not coincide with any of the individual rankings. Given the restriction $\alpha=0$, for any $a_{1},a_{2}\in\Re$ the function $e(x) = u_{0}(x) - a_{1}u_{1}(x)-a_{2}u_{2}(x)$ has $\omega_{e}(X)\geq 1$, so the $\epsilon$-Pareto Indifference condition fails for any $\epsilon\in [0,1)$. In particular, Pareto Indifference is also violated. However, for $\alpha\in (0,1)$, where Pareto Indifference also fails, we have for $e(x) = u_{0}(x) - \alpha u_{2}(x)$ that $\omega_{e}(X)\leq 1-\alpha$, so we obtain an approximate utilitarian aggregator given $(\epsilon,\W_{+})$ if $1-\alpha\leq \epsilon<1$. Therefore, either a nontrivial approximate utilitarian aggregation is impossible (the case $\alpha=0$), or there exists one for suitable choices of $\epsilon$ when $\alpha\in (0,1)$. In the latter case, the $\epsilon$-Semistrong Pareto axiom holds.\myqed
\end{example}


\subsection{The case with positive weights}


Consider the aggregation problem with approximate solutions in the set $\W_{++}$. The following axiom is a variation of the Strong Pareto principle that ensures no Pareto weight in Theorem \ref{theorem: semistrong without compactness} is zero when utilitarian aggregation is exact.

\begin{definition}[$\epsilon$-Strong Pareto]\label{definition: epsilon strong simple}
	The pair $(u_{0},\U)$ satisfies $\epsilon$-Strong Pareto when for all $x,y\in X$: if $u_{i}(x)\geq u_{i}(y)$ for all $i=1,\dots,N $, then $u_{0}(x)\geq u_{0}(y)-\epsilon$; and, if in addition $u_{i}(x)>u_{i}(y)$ for some $i$, then $u_{0}(x)>u_{0}(y)-\epsilon$.
\end{definition}

Note that $\epsilon$-Strong Pareto strengthens the conclusion of the $\epsilon$-Semistrong Pareto axiom. Specifically, $\epsilon$-Strong Pareto is equivalent to combining $\epsilon$-Pareto Indifference with the condition that $u_{0}(x)>u_{0}(y)-\epsilon$ whenever $u_{i}(x)\geq u_{i}(y)$ for all $i=1,\dots,N$,  with strict inequality for at least one $i$.

However, the weaker form of Strong Pareto in Definition \ref{definition: epsilon strong simple} is necessary but insufficient to ensure that the Pareto weights in the approximate utilitarian aggregation are strictly positive. As the next example suggests, to obtain an approximate form of aggregation with strictly positive Pareto weights, we shall require a stronger version of $\epsilon$-Strong Pareto.

\begin{example}\label{example: fails positive weights}
	Consider $N=1$, $X = \{(x_{1},x_{2})\in\Re^{2}:x_{1}^{2}+x_{2}^{2}\leq 1, x_{1}\geq 0,x_{2}\geq 0\}$, and $\epsilon=1$. For $x = (x_{1},x_{2})$ the utility functions are given by the expressions $u_{0}(x) = -x_{1}$ and $u_{1}(x)=x_{2}$. They are both affine mappings on the convex set $X$. Since $\omega_{u_{0}}(X)= 1$ we certainly have $u_{0}(x)\geq u_{0}(y)-1$ when $u_{1}(x)\geq u_{1}(y)$. And if $u_{1}(x)>u_{1}(y)$ then $x_{2}>y_{2}$. When $u_{0}(x)>u_{0}(y)-1$ is false, we have that $u_{0}(x) = u_{0}(y)-1$ because $\omega_{u_{0}}(X)= 1$. Therefore $x_{1} = y_{1}+1\geq 1$, and thus $0=x_{2}>y_{2}$, which is impossible. At the same time, there is no $a_{1}>0$ for which the function $e= u_{0}-a_{1}u_{1}$ has oscillation $\omega_{e}(X)\leq 1$.  In fact, with the choice of $x = \left(\sqrt{\frac{1}{1+a_{1}^{2}}}, \sqrt{\frac{a_{1}^{2}}{1+a_{1}^{2}}} \right)$ and $y = (0,0)$ in $X$, we obtain that $\omega_{e}(X)\geq x_{1}+a_{1}x_{2} - y_{1} - a_{1}y_{2}=\sqrt{1+a_{1}^{2}}>1$ when $a_{1}>0$. \myqed
\end{example}

When we frame the solution to the aggregation problem under Strong Pareto as characterizing Pareto optimality via scalarization over a convex set formed with profiles of utility differences, the previous example mirrors a standard case showing that Pareto optimal points are not necessarily ``proper.''\footnote{Recall that Pareto optimal points, or simply efficient points, in a subset of  $\Re^{M}$ cannot be strictly improved in one coordinate without reducing at least one other coordinate. In contrast, properly efficient points are efficient points for which a fixed multiple of the maximum losses in other coordinates bounds gains in (any) one coordinate (see, e.g., \cite{jahn2009}).} In this context, efficient points that are not proper cannot be viewed as solutions to the maximization of a linear function with strictly positive coefficients, which reflects a similar limitation encountered when seeking approximate utilitarian aggregation rules with strictly positive weights. Accordingly, the necessary modifications to the $\epsilon$-Strong Pareto axiom can be made using the concept of proper efficiency. However, the full extent of this modification is not required in our setting, due to the additional structure provided by the vN-M utilities and our formulation of the $\epsilon$-Pareto axioms. To be precise, our alternative version of $\epsilon$-Strong Pareto combines the $\epsilon$-Semistrong Pareto axiom with conditions reminiscent of those in \citeauthor{geoffrion1968}'s (\citeyear{geoffrion1968}) notion of proper efficiency. This version of the Pareto axiom is defined below. 

\begin{definition}[Sequential $\epsilon$-Strong Pareto]\label{definition: sequential epsilon strong}
	The pair $(u_{0},\U)$ satisfies Sequential $\epsilon$-Strong Pareto when the following two conditions hold.
	\begin{itemize}
		\item[(a)] For no $x,y\in X$ we have at the same time $u_{0}(y)-u_{0}(x)-\epsilon>0$ and ${u_{i}(x)-u_{i}(y)}\geq 0$ for all $i\in\{1,\dots,N\}$.
		\item[(b)] For no pair of sequences $(x_{n})$ and $(y_{n})$ in $X$ there exists $i\in \{1,\dots,N\}$ such that $u_{i}(x_{n})-u_{i}(y_{n})>0$ for all $n$, $\liminf_{n\to\infty}\frac{u_{j}(x_{n})-u_{j}(y_{n})}{u_{i}(x_{n})-u_{i}(y_{n})}\geq 0$ for all $j\neq i$ with $j\in\{1,\dots,N\}$, and $\liminf_{n\to\infty}\frac{u_{0}(y_{n})-u_{0}(x_{n})-\epsilon}{u_{i}(x_{n})-u_{i}(y_{n})}\geq 0$.
	\end{itemize}
\end{definition}

The Sequential $\epsilon$-Strong Pareto axiom builds on $\epsilon$-Semistrong Pareto by adding the requirement described in part (b) of Definition \ref{definition: sequential epsilon strong}. When applied to constant sequences, part (b) reduces to the additional condition that distinguishes $\epsilon$-Strong Pareto from $\epsilon$-Semistrong Pareto. Notably, the Sequential $\epsilon$-Strong Pareto axiom prevents a pair of alternatives $x$ and $y$ from being approached by sequences of alternatives $(x_{n})$ and $(y_{n})$ such that, for some individual $i$, $x_{n}\succ_{i}y_{n}$, while for all other individuals $j\neq i$ and the DM, the preferences $x\pref_{j}y$ (for $j\neq 0$) hold along with the failure, in terms of utility, of the strict form of the approximate optimality condition in (\ref{equation: nearly preferable}) \textit{in the limit}. 

Our Definition \ref{definition: sequential epsilon strong} is even stronger than previously described because part (b) remains valid when alternatives are replaced with their corresponding utility values, without requiring the alternatives themselves to converge. To make this more concrete, we assume that all individuals and the DM have a best and a worst alternative in $X$. In this case, denote by $\xh_{j}$ and $\xl_{j}$ the best and worst alternatives, respectively, for $j=0,1,\dots,N$. Suppose also that $\xh_{j}\succ_{j}\xl_{j}$. Then for each alternative $x$, there is a unique $\lambda_{j,x}\in [0,1]$ with the property that $x\sim_{j}\lambda_{j,x}\xh_{j}+(1-\lambda_{j,x})\xl_{j}$. When the parameter $\epsilon$ is chosen so that $\epsilon = \lambda_{0,x^{\ast}}-\lambda_{0,x_{\ast}}<1$ for some pair of alternatives $x^{\ast}$ and $x_{\ast}$ with $\xh_{0}\succ_{0}x^{\ast}\succ_{0}x_{\ast}\succ_{0}\xl_{0}$, the inequalities involving the limit inferior in part (b) of Definition \ref{definition: sequential epsilon strong} become
\begin{align}\label{equation: liminf using lambdas}
	\liminf_{n\to\infty}\frac{\lambda_{j,x_{n}}-\lambda_{j,y_{n}}}{\lambda_{i,x_{n}}-\lambda_{i,y_{n}}}\geq 0\quad\text{and}\quad \liminf_{n\to\infty}\frac{\lambda_{0,y_{n}}-\lambda_{0,x_{n}} - (\lambda_{0,x^{\ast}}-\lambda_{0,x_{\ast}})}{\lambda_{i,x_{n}}-\lambda_{i,y_{n}}}\geq 0.
\end{align} 
Although the Sequential $\epsilon$-Strong Pareto axiom can be expressed in terms of empirically observable objects, we note that falsifying it using the inequalities in (\ref{equation: liminf using lambdas}) typically requires verifying infinitely many conditions, which is infeasible.

Before presenting our general aggregation result with positive weights, we first observe that the setting with vN-M utilities in Example \ref{example: fails positive weights} does not satisfy the Sequential $\epsilon$-Strong Pareto axiom. This is shown below.

\begin{examplec}{\ref{example: fails positive weights} (continued)}
	The Sequential $\epsilon$-Strong Pareto axiom fails in this example. To see this, take the sequences $(x_{n})$ and $(y_{n})$ where $x_{n} = \left(\sqrt{1-\frac{1}{n^{2}}},\frac{1}{n}\right)$ and $y_{n} = \left(0,0\right)$. Then $u_{1}(x_{n})-u_{1}(y_{n})=\frac{1}{n}$ and $u_{0}(y_{n}) -u_{0}(x_{n})-\epsilon =  \sqrt{1-\frac{1}{n^{2}}}-1$. Consequently,
	\begin{align*}
		\lim_{n\to\infty} \frac{u_{0}(y_{n}) -u_{0}(x_{n})-\epsilon}{u_{1}(x_{n})-u_{1}(y_{n})} = \lim_{n\to\infty} \sqrt{n^{2}-1}-n=0, 
	\end{align*}
	thus violating the Sequential $\epsilon$-Strong Pareto condition. \myqed
\end{examplec}

As noted above, the failure of Sequential $\epsilon$-Strong Pareto in the previous example parallels the canonical example that illustrates the distinction between efficient and properly efficient points. Examples of this kind appear, for instance, in \cite{borwein1977,borwein1980}, \citet[p.~37]{sawaragi1985}, and \citet[p.~288]{jahn2009}. More recently, \cite{acciaioetal2022} presented a similar example in a paper in which the standard no-arbitrage conditions are replaced by weaker ones, akin to the weaker Pareto axioms discussed in this paper. The connection between their paper and ours is partly explained by the relationship between the utilitarian aggregation problem and asset pricing under no-arbitrage, as reported in \cite{turunenred1999}, due to their similar mathematical structure. A common unifying idea is that both problems can be viewed as characterizing a Pareto optimal point via scalarization. As we shall see below in Theorem \ref{theorem: aggregation strong polyhedral}, with finitely many individuals, examples of this form are only possible when the set in which we aim to characterize the efficient points is not the convex hull of finitely many points. 

The following theorem characterizes approximate utilitarian aggregation with positive weights. Like the previous results, it relies on a suitable application of the minimax theorem. 

\begin{theorem}\label{theorem: aggregation with sequential epsilon strong}
Let $\epsilon\geq 0$, and suppose that  $(X,u_{0},\U)$ is a setting with vN-M utilities. The following statements are equivalent.
	\begin{itemize}
		\item[(i)] The pair $(u_{0},\U)$ satisfies Sequential $\epsilon$-Strong Pareto.
		\item[(ii)] The utility function $u_{0}$ is approximately utilitarian given $(\epsilon,\W_{++})$.
	\end{itemize}
\end{theorem} 

As anticipated, the next theorem identifies a case in which the specific structure of the set $X$ and the utility functions ensure that $\epsilon$-Strong Pareto and its sequential version are equivalent. This corresponds to a setting in which the set of alternatives is an abstract set of lotteries over a finite number of prizes.

\begin{theorem}\label{theorem: aggregation strong polyhedral}
	Let $\epsilon\geq 0$, and suppose that  $(X,u_{0},\U)$ is a setting with vN-M utilities where $X$ is the convex hull of finitely many points. The following statements are equivalent.
	\begin{itemize}
		\item[(i)] The pair $(u_{0},\U)$ satisfies $\epsilon$-Strong Pareto.
		\item[(ii)] The utility function $u_{0}$ is approximately utilitarian given $(\epsilon,\W_{++})$. 
	\end{itemize}
\end{theorem}

The proof of Theorem \ref{theorem: aggregation strong polyhedral}  leverages the special structure of $X$ to ensure that the set of profiles of utility differences, as defined in equation (\ref{equation: define z epsilon}) in Appendix \ref{appendix}, constitutes a subset of the standard Euclidean space of dimension $N+1$ which is also the convex hull of finitely many points. This property is key to establishing that the $\epsilon$-Strong Pareto axiom implies the Sequential $\epsilon$-Strong Pareto condition.  A similar result, based on the polyhedrality of sets, demonstrates that Pareto efficiency and proper efficiency coincide in sets with this structure (see, e.g., Theorem 3.4.7 in \cite{sawaragi1985}). More generally, solution sets of systems of finitely many (weak) linear inequalities in finite-dimensional spaces have the property that their efficient points are also properly efficient relative to the usual ordering cone. In the proof of Theorem \ref{theorem: aggregation strong polyhedral}, we exploit this type of structure precisely, arising from a set formed by convex combinations of profiles of utility differences and a vector whose first coordinate contains the parameter $\epsilon$.

As a final observation, the following example, adapted from \citet[pp.\ 274-275]{weymark1991}, shows that our version of $\epsilon$-Strong Pareto can sometimes be overly permissive in allowing approximate aggregation. Specifically, it suggests a case where the Strong Pareto axiom fails, yet its weaker version holds for every choice of $\epsilon>0$.

\begin{example}
Suppose that $N=2$, and  $X = \{(x_{1},x_{2})\in\Re^{2}:x_{1}+x_{2}\leq 1, x_{1}\geq 0,x_{2}\geq 0\}$. For $x=(x_{1},x_{2})$ the utility functions are given by $u_{0}(x)=u_{1}(x)=x_{1}$ and $u_{2}(x)=-x_{1}-x_{2}$. When $\epsilon=0$, the standard Semistrong Pareto axiom is satisfied in this setting, but Strong Pareto is not. The failure of Strong Pareto arises from the fact that $u_{1}(x) = u_{1}(y)$ and $u_{2}(x)>u_{2}(y)$ imply that $u_{0}(x)=u_{0}(y)$. However, for any $\epsilon>0$ we trivially have $u_{0}(x)>u_{0}(y)-\epsilon$, so the $\epsilon$-Strong Pareto axiom is satisfied. Indeed, choosing weights $a_{1} = 1$ and $0<a_{2}\leq\epsilon$, for $e = u_{0}-a_{1}u_{1}-a_{2}u_{2}$ we obtain that $e(x) - e(y) = a_{2}(x_{1}+x_{2} - y_{1}-y_{2})\leq a_{2}(x_{1}+x_{2})\leq \epsilon$. Hence $\omega_{e}(X)\leq \epsilon$,  which shows that $u_{0}$ is approximately utilitarian given $(\epsilon,\W_{++})$.\myqed  
\end{example}

\subsection{Two more examples}\label{section SEU examples}


Suppose preferences are modeled in the Anscombe-Aumann framework, as described in Example \ref{example SEU 1}. As shown by  \cite{seidenfeld1989} and \cite{mongin1995,mongin1998}, the standard Pareto conditions can be overly restrictive in this setting. In particular, the Pareto conditions may rule out non-dictatorial aggregation rules when individual utilities exhibit a rich structure. 

Consider the case of a finite set $S$ of states. Suppose that, for each $j = 0, 1, \dots, N$, preferences $\pref_{j}$ are represented by a utility function of the form
\begin{align*}
	u_{j}(f) = \sum_{s\in S}\mu_{j}(s)v_{j}(f(s)),
\end{align*}
where  $\mu_{j}$ is a subjective probability and $v_{j}\colon C\to\Re$ is an affine state-independent utility function representing tastes. Proposition 4 of \cite{mongin1998}  shows that, under the Semistrong Pareto condition, if the sets  $\{\mu_{1},\dots,\mu_{N}\}\subseteq \Re^{S}$ and $\{v_{1},\dots,v_{N},\indic_{C}\}\subseteq\Re^{C}$ are each linearly independent, then the utilitarian aggregation must result in both a utility dictator and a probability dictator.

If, instead of adopting an exact form of utilitarian aggregation, we consider the approximate form developed in this paper, it is not difficult to show that the assumptions of \cite{mongin1998} are insufficient to guarantee the existence of a probability dictator. More precisely, if individuals' tastes are sufficiently close, a non-dictatorial approximate utilitarian aggregation is possible under the same linear independence assumptions as in \cite{mongin1998}.
To illustrate, suppose that for each individual $i\in \{2,\dots,N\}$, preferences over the set $C$ of consequences are nearly represented by the same utility $v_{1}$ that represents the tastes of individual $1$. Specifically, assume that $\norm{v_{1} - v_{i}}_{\infty}\leq \frac{\epsilon}{2}$. These constraints on the $v_{i}$'s do not preclude the set $\{v_{1},\dots,v_{N},\indic_{C}\}$ from being linearly independent. An example showing this point is given below.

\begin{example}\label{example: first additional SEU}
	Let $C$ be the set of all lotteries over three deterministic prizes. Suppose that $S$ is also a three-element set and $N=2$. Since both $v_{i}$ and $\mu_{i}$ can be identified with vectors in $\Re^{3}$, we assume that $v_{1} = \left (\frac{3\epsilon}{2} ,\frac{\epsilon}{4},1\right)$,  $v_{2} = \left (\epsilon ,-\frac{\epsilon}{4},1\right)$, $\mu_{1}=\left (\frac{1}{2},\frac{1}{4},\frac{1}{4}\right)$ and $\mu_{2} = \left(\frac{1}{4},\frac{1}{4},\frac{1}{2}\right)$. Then $\norm{v_{2}-v_{1}}_{\infty}\leq\frac{\epsilon}{2}$, and the linear independence assumption for tastes and probabilities is satisfied. We note that the DM's utility as defined by
	\begin{align*}
		u_{0}(f) = \frac{1}{2}u_{1}(f) + \frac{1}{2}u_{2}(f)
	\end{align*} 
	can also be expressed as the sum of a subjective expected utility representation with tastes $v_{1}$ and prior $\frac{1}{2}\mu_{1}+\frac{1}{2}\mu_{2}$, plus a residual term of norm at most $\frac{\epsilon}{2}$. We still retain the existence of a utility dictator in the representation of $u_{0}$, but there need not be a probability dictator.\myqed
\end{example}

Beyond Example \ref{example: first additional SEU}, suppose the individuals' tastes are sufficiently similar, in the sense that $\norm{v_{i}-v_{1}}_{\infty}\leq\frac{\epsilon}{2}$, and consider a list of weights $\lambda_{1},\dots,\lambda_{N}\geq 0$ with $\sum_{i=1}^{N}\lambda_{i}=1$. Then one can verify that the function $f\mapsto \sum_{s\in S}\mu(s)v_{1}(f(s))$, where $\mu = \sum_{i=1}^{N}\lambda_{i}\mu_{i}$, is $\frac{\epsilon}{2}$-close to $u_{0} = \sum_{i=1}^{N}\lambda_{i}u_{i}$. In other words, if the linear independence of the relevant sets reflects a rich structure on the sets of tastes and priors, then Mongin's conclusion that a probability dictator is necessary does not carry over to our approximately utilitarian setting when tastes are sufficiently close. This shows that a non-degenerate linear pooling of probabilistic opinions remains possible -- albeit in an approximate form -- even without complete agreement on tastes.

Now consider the case of Example \ref{example SEU 2}, where the more general set of consequences $C$ contains at least one pair of elements that are strictly and unanimously ranked. The set of states, $S$, is not assumed to be finite, and each prior is a non-atomic probability measure. Recall that in this framework, two important subsets of the set of alternatives give rise to distinct settings with vN-M utilities. One such subset is the set of all lotteries, that is, the set of all acts $f$ that induce a common finitely supported probability distribution $p$ on the set $C$ of consequences. In this context, our version of the Semistrong Pareto condition, when restricted to the domain $\mathcal{L}(C)$ of such lotteries, corresponds to a version of the Lottery Pareto axiom introduced by \cite{gilboa2004}.

\begin{definition}[$\epsilon$-Lottery Pareto]\label{definition: epsilon lottery}
	We say that the pair $(u_{0},\U)$ satisfies the $\epsilon$-Lottery Pareto axiom when for all $p,q\in\mathcal L(C)$: if $u_{i}(p)\geq u_{i}(q)$ for all $i=1,\dots,N$, then $u_{0}(p)\geq u_{0}(q)-\epsilon$.	
\end{definition}

As in the original condition, our $\epsilon$-Lottery Pareto principle applies to situations where the individuals and the DM agree on the likelihoods of the events. It can thus be viewed as the $\epsilon$-Semistrong Pareto condition restricted to the subset $\mathcal L(C)$ of lotteries within the broader domain of Savage acts mentioned in Example \ref{example SEU 2}.

Given our framework of approximately utilitarian aggregation, it is also important to establish conditions under which the prior associated with the DM's utility function $u_{0}$ is approximately a linear combination of the individual priors. In \cite{gilboa2004}, the authors show that their Restricted Pareto condition is sufficient for belief aggregation. In our setting, we adopt a version of the Likelihood Pareto condition introduced in \cite{alon2016}. To this end, recall the expression in (\ref{equation: sigmau defined}), where $\Sigma_{u}$ denotes the set of all events whose probabilities are agreed upon by both the DM and the individuals. For any $F\in \Sigma_{u}$ and $\epsilon\in[0,1]$, we define $F_{\epsilon} \subseteq F$ as an element of $\Sigma_{S}$ such that $P_{0}(F_{\epsilon})=\epsilon$ if $\epsilon\leq  P_{0}(F)$, and $F_{\epsilon}=F$ otherwise. Since $P_{0}$ is a non-atomic probability measure, such an event $F_{\epsilon}$ is always well-defined.

Our next approximate Pareto dominance condition uses the events in $\Sigma_{u}$, along with the existence of unanimously ranked consequences $c^{\ast}$ and $c_{\ast}$, to elicit bounds on the likelihood of each event in $\Sigma_{S}$ under $P_{0}$. It plays an essential role in determining the approximate aggregation of probabilities.  

\begin{definition}[$\epsilon$-Likelihood Dominance]\label{definition: epsilon likelihood}
	We say that the pair $(u_{0},\U )$ satisfies $\epsilon$-Likelihood Dominance when for all $E\in\Sigma_{S}$ and $F\in\Sigma_{u}$: if $u_{i}(c^{\ast}Ec_{\ast})\geq u_{i}(c^{\ast}Fc_{\ast})$ for all $i=1,\dots,N$, then $u_{0}(c^{\ast}Ec_{\ast})\geq u_{0}(c^{\ast}Gc_{\ast})$ for $G=F\setminus F_{\epsilon}$.
\end{definition}

Using Definitions \ref{definition: epsilon lottery} and \ref{definition: epsilon likelihood}, we can now state our separate and approximate utilitarian aggregation results for tastes and probabilities.

\begin{proposition}\label{proposition: SEU epsilon}
	Let $\epsilon_{1}\geq 0$ and $\epsilon_{2}\in[0,1]$. Assume that each utility function is of the form described in Example \ref{example SEU 2}. The following statements are equivalent.
	\begin{enumerate}
		\item[(i)] The pair $(u_{0},\U)$ satisfies $\epsilon_{1}$-Lottery Pareto and $\frac{\epsilon_{2}}{2}$-Likelihood Dominance.
		\item[(ii)] There exist real numbers $a_{1},\dots,a_{N}\geq 0$ and $b$, and $\lambda_{1},\dots,\lambda_{N}\geq 0$ with $\sum_{i=1}^{N}\lambda_{i}=1$ such that, for $w_{0}(c) =\sum_{i=1}^{N}a_{i}v_{i}(c)+b $,  $v_{0} = w_{0}+r$ for some function $r\colon C\to \Re$ with $\norm{r}_{\infty}\leq \frac{\epsilon_{1}}{2} $, and $P_{0} = \sum_{i=1}^{N}\lambda_{i}P_{i}+R$ for some signed measure $R$ with $\norm{R}_{1}\leq \epsilon_{2}$.
	\end{enumerate}
\end{proposition}

\begin{remark}
	In the proof of Proposition \ref{proposition: SEU epsilon}, we employ Theorem \ref{theorem: semistrong without compactness} to derive an approximately utilitarian aggregation of tastes. Similar results hold under alternative conditions on the Pareto weights, provided the $\epsilon$-Lottery Pareto axiom is adjusted accordingly, in line with Theorems \ref{theorem: aggregation finite indifference} and \ref{theorem: aggregation with sequential epsilon strong}. To aggregate individual probabilities over the subdomain of Savage acts with agreed-upon outcomes, we identify a setting with vN-M utilities in which our main theorems also apply. In fact, by applying Theorem \ref{theorem: semistrong without compactness} with a suitable modification of $\epsilon$-Likelihood Dominance, we could have achieved an aggregation of the form $P_{0} = \sum_{i=1}^{N}\lambda_{i}P_{i}+R$ with $\lambda_{i}\geq 0$ and $\norm{R}_{1}\leq \epsilon_{2}$. While such a result would retain the structure of approximately linear aggregation, we could not prove, in this case, that the normalization $\sum_{i=1}^{N}\lambda_{i}=1$ is possible. The additional work in the proof of Proposition \ref{proposition: SEU epsilon} reflects our attempt to obtain such a normalization using the condition $P_{0}(E)\geq \min_{1\leq i\leq N}P_{i}(E)-\frac{\epsilon_{2}}{2}$, as is done in Proposition 7 in \cite{nascimento2024} in a related setting with additional assumptions on the set of states. 
	\myqed
\end{remark}


\section{Concluding remarks}\label{section: conclude}

In this paper, we characterized approximate forms of utilitarian aggregation for finitely many vN-M utilities. Our approximate versions of the exact aggregation results revealed a form of stability in the aggregation problem, in the sense that small violations of the Pareto principles lead to aggregation rules that remain approximately utilitarian. This is related to an old question posed by \citet[p.~63]{ulam1960} regarding the stability of certain functional equations. Broadly speaking, Ulam's question asks whether a problem that deviates slightly from another still admits a solution that remains close to that of the original problem.\footnote{See \cite{hyers1998} for an account of the early results in the literature on the stability of functional equations.}  In analyzing the sensitivity of the utilitarian aggregation rules to slight changes in the Pareto axioms, we showed that the same parameter $\epsilon$  that bounds violations of the Pareto unanimity conditions also bounds the distance between the DM's utility function and a utilitarian aggregation function.  We therefore answered Ulam's question in the affirmative in the context of utilitarian aggregation of vN-M utilities.\footnote{In related work, and also in connection with a form of stability in Harsanyi’s aggregation theorem, \cite{mccarthy2020} showed that, under smoothness assumptions, violations of the Independence and Pareto axioms can still yield a social preorder that ``locally'' resembles a utilitarian aggregator. I thank David McCarthy for pointing out this connection.}

When interpreting the results of this paper as a quantitative extension of classic aggregation theorems, we recall that Theorem \ref{theorem: duality V} established a duality formula under additional assumptions on the setting with vN-M utilities. While these assumptions hold in many significant applications, in the more general settings of Theorems \ref{theorem: semistrong without compactness}, \ref{theorem: aggregation finite indifference}, and \ref{theorem: aggregation with sequential epsilon strong}, we provided an upper bound on the distance between the DM's utility function and a class of utilitarian aggregators. A related duality result and upper bounds were obtained in \cite{nascimento2024}, where the maximum sure gains with normalized bets characterized the distance of a probability measure from a set of probabilities possessing certain properties. As in the present paper, the minimax theorem was also a central tool in deriving the duality formula in \cite{nascimento2024}.\footnote{To gain some intuition for using the minimax theorem, we refer to \cite{nau1991}, who discuss several connections between linear programming duality and forms of no-arbitrage. See also \cite{nau2025} for a more comprehensive treatment.} Concerning this paper's approximate utilitarian aggregation results, we suspect that the duality result in Theorem \ref{theorem: duality V} can be extended more generally to a setting with vN-M utilities without additional topological assumptions. Nevertheless, the mathematical problem likely parallels the arguments in the proof of Theorem \ref{theorem: semistrong without compactness}, and as such, a more general duality result lies beyond the scope of this paper.

At the same time, unlike the early proofs of Harsanyi's aggregation theorem in \cite{domotor1979}, \cite{fishburn1984}, \cite{border1985}, \cite{weymark1991}, and \cite{demeyer1995}, our results rely on an application of the minimax theorem. Our approach builds on the key observation that the standard Pareto axioms can be framed in terms of efficiency in the space of profiles of utility differences, using a suitably defined domination structure. For instance, in the case of a single individual ($N=1$), the Strong Pareto condition is equivalent to the statement that the point $(0,0)$ in $\Re^{2}$ is a Pareto efficient point (under the usual order) of the symmetric and convex set $Z = \{ (u_{0}(y) - u_{0}(x), u_{1}(x)-u_{1}(y)): x,y\in X \}$. It is also not difficult to see that $(0,0)$ remains an efficient point of the vector space spanned by $Z$, thus allowing the standard utilitarian aggregation results to be derived via scalarization techniques for finite-dimensional vector optimization problems under polyhedrality conditions.\footnote{\label{footnote: thms alternative}In the case of a space of lotteries with finitely many prizes, our results can also be derived with (non-homogeneous) theorems of the alternative. In this regard, they are similar to early proofs involving theorems of the alternative. These are found in \cite{selinger1986} and \cite{weymark1994}.} However, the theorems in this paper cannot rely on such a simplification, as the relevant efficient point contains the parameter $\epsilon$ in its first coordinate. We applied the minimax theorem to address this challenge and developed a unified proof method that accounts for the domination structures implied by our versions of the Pareto axioms.

We also note that our analysis was restricted to the case of a finite set of utilities $\U$, as defined in (\ref{equation: set U finite}). The methods employed in this paper do not appear to extend to settings with infinitely many individuals. We suspect that our approach can be extended to the setting of \cite{zhou1997}, where the set $\U$ need not be finite.\footnote{For settings that also deal with an aggregation problem with an infinite number of preference orderings, see also \cite{dananetal2013, dananetal2016}. Closely related to Zhou's single-profile setting, we also refer the reader to \cite{askoura2018, askoura2021} and \cite{nascimento2011}.} At least in the case of our $\epsilon$-Semistrong Pareto axiom, and in a setting involving lotteries over a compact metric space, methods of best approximation in normed spaces (e.g., \cite{singer1970}) could be employed to establish an approximate utilitarian aggregation theorem.\footnote{It seems that the same methods of best approximation could also be used in a setting with a finite $\U$ and with lotteries over finitely many prizes to establish uniqueness results under the Independent Prospects condition of \cite{weymark1991}. For instance, under the $\epsilon$-Pareto Indifference axiom, the utilitarian aggregation function in $\W$ that is closest to $u_{0}$ is apparently unique with the suitable choice of the Euclidean norm instead of the maximum norm used in this paper.} Achieving this would require additional assumptions on $\U$, such as compactness, to properly frame the problem as finding the nearest point in a closed convex set to the utility function $u_{0}$. 

Finally, requiring a sufficiently strong preference for $x$ over $y$ according to the utilities in $\U$ to induce the DM to weakly prefer $x$ to $y$ can be viewed as a dual version of the $\epsilon$-Semistrong Pareto results established in this paper. Specifically, $x\pref_{0}y$ when the difference $u_{i}(x)-u_{i}(y)$ is at least a specified nonnegative value $\epsilon_{i}$ for every individual $i$. When the set of alternatives $X$ consists of lotteries over finitely many deterministic outcomes, the condition
\begin{align}\label{equation: Semistrong dual}
	u_{0}(x)\geq u_{0}(y) \quad \text{ when }\quad u_{i}(x)\geq u_{i}(y)+\epsilon_{i} \text{ for all }i=1,\dots,N,
\end{align}
implies, via a non-homogeneous version of Farkas' lemma (as mentioned in Footnote \ref{footnote: thms alternative}), the existence of weights $a_{i}\geq 0$, for $i=1,\dots,N$, such that $u_{0} = \sum_{i=1}^{N}a_{i}u_{i}+e$, where
\begin{align}\label{equation: osc ui epsiloni}
	\omega_{e}(X)\leq \sum_{i=1}^{N}a_{i}\epsilon_{i}.
\end{align}
The relative appeal of our results based on $\epsilon$-Semistrong Pareto, as opposed to condition (\ref{equation: Semistrong dual}), is that the bound in (\ref{equation: osc ui epsiloni}) requires interpersonal comparisons of the thresholds $\epsilon_{i}$. As a result, the bound on the distance between $u_{0}$ and a utilitarian aggregator also depends on the Pareto weights. The approach in this paper sidesteps this issue by adopting the DM's perspective from the outset.

%
%
%



\appendix
\counterwithin{lemma}{subsection}

\section{Proofs}\label{appendix}


\subsection{Proof of Proposition \ref{proposition: equivalent forms of approximation}}

 The two lemmas below make precise the geometric intuition given before the statement of Proposition \ref{proposition: equivalent forms of approximation}. The fact that (i) implies (ii) is a consequence of Lemma \ref{lemma: auxiliary geometric}, while the converse implication follows from Lemma \ref{lemma: auxiliary 2}.
  
 \begin{lemma}\label{lemma: auxiliary geometric}
  Suppose that $D$ is a nonempty set, and that $h_{j}\colon D\to \Re $, for $j=0,1,\dots,N$ are functions with the property that, for some $a_{1},\dots,a_{N}\in \Re $, the function
  \[
  h = h_{0}-\sum_{i=1}^{N}a_{i}h_{i}
  \]
 satisfies $\omega_{h}(D)<\infty$ (equivalently, $h$ is bounded). Then there exists $b\in \Re$ such that
 \[
 \norm{h-b\indic_{D}}_{\infty} = \frac{\omega_{h}(D)}{2}.
 \]
  \begin{proof}
  	Since $h$ is bounded, both $\sup_{d\in D}h(d)$ and $\inf_{d\in D}h(d)$ exist, and  we define
  	\begin{align*}
 	b = \frac{\sup_{d\in D}h(d) +\inf_{d\in D}h(d)}{2}.
 	 \end{align*}
 	Then
 	\begin{align*}
 	h(d)-b\leq \sup_{d\in D} h(d)- b=\frac{\omega_{h}(D)}{2}.
     \end{align*}
     For a similar reason, 
\begin{align*}
	h(d)  - b \geq  \inf_{d\in D}h(d) - b =  - \frac{\omega_{h}(D)}{2}.
\end{align*}
 Therefore, $\norm{h-b\indic_{D}}_{\infty} \leq  \frac{\omega_{h}(D)}{2}$. If the inequality is strict, by definition of supremum there are $d,d^{\prime}\in D$ with $\norm{h-b\indic_{D}}_{\infty}< \frac{h(d)-h(d^{\prime})}{2} \leq\frac{\omega_{h}(D)}{2} $, which is impossible since $ \frac{h(d)-h(d^{\prime})}{2} = \frac{h(d) - b -(h(d^{\prime})-b)}{2}\leq \norm{h-b\indic_{D}}_{\infty}$.
  \end{proof} 	
 \end{lemma}
 
\begin{lemma}\label{lemma: auxiliary 2}
  Suppose that $D$ is a nonempty set, and that $h_{j}\colon D\to \Re $, for $j=0,1,\dots,N$ are functions such that, for some $a_{1},\dots,a_{N},b\in\Re$, the function
  \[
  	h_{0}-\sum_{i=1}^{N}a_{i}h_{i} - b\indic_{D}
  \]
   is bounded. Define $h =h_{0}-\sum_{i=1}^{N}a_{i}h_{i} $. Then 
   \[
   \omega_{h}(D)\leq 2\norm{h_{0}-\sum_{i=1}^{N}a_{i}h_{i} - b\indic_{D}}_{\infty}.
   \]
	\begin{proof}
		The oscillation of $h$ is
		\begin{align*}
			\sup_{d\in D}h(d) -\inf_{d^{\prime}\in D}h(d^{\prime})=\sup_{d\in D}\left(h_{0}(d)-\sum_{i=1}^{N}a_{i}h_{i}(d)\right) -b - \inf_{d^{\prime}\in D}\left(h_{0}(d^{\prime})-\sum_{i=1}^{N}a_{i}h_{i}(d^{\prime})\right)+b,
		\end{align*}
		which is not greater than $2\norm{h_{0}-\sum_{i=1}^{N}a_{i}h_{i} - b\indic_{D}}_{\infty}$.
	\end{proof}
\end{lemma}  

Finally, the additional statement is an immediate consequence of the equivalence just established without reference to the sign of the coefficients $a_{i}$.


\subsection{Proof of Theorem \ref{theorem: duality V}}

The proof rests on the following lemma.

\begin{lemma}\label{lemma: minimax duality}
	Suppose that  $(X,u_{0},\U)$ is a setting with continuous vN-M utilities and a compact domain, and that $\V$ is the subset of $X\times X$ defined in (\ref{equation: V defined}). Then, the function $\eta\colon X\times X\times \Re_{+}^{N}\to \Re$ defined by
	\begin{align}\label{equation: eta defined}
		\eta(x,y,\bm a) = u_{0}(y)-u_{0}(x)+ \sum_{i=1}^{N}a_{i}[u_{i}(x)-u_{i}(y)]
	\end{align}
	has the minimax property
	\begin{align}\label{equation: minimax property}
		\max_{(x,y)\in X\times X}\inf_{\bm  a\in\Re^{N}_{+}}\eta(x,y,\bm a) =\min_{\bm  a\in\Re^{N}_{+}}\max_{(x,y)\in X\times X} \eta(x,y,\bm a).
	\end{align}
	\begin{proof}
		First, if all the functions $u_{i}$ are constant, we must have that $\eta(x,y,\bm a)=u_{0}(y)-u_{0}(x)$, so the minimax property (\ref{equation: minimax property}) is trivially satisfied, and for $\bm a = 0$ we have  $e =u_{0}$, thus showing that (\ref{equation: duality V}) holds. Therefore, we assume that at least one of the $u_{i}$'s is not a constant function for the relevant case.

		We endow the Cartesian product $X\times X$ with the product topology, which makes this set compact. Define the functions $U_{0},U_{i}\colon X\times X\to\Re$ so that
			\begin{align}\label{equation: functions U0 Ui defined}
				U_{0}(x,y)= -[u_{0}(x)-u_{0}(y)] \quad\text{and}\quad U_{i}(x,y) = u_{i}(x) -u_{i}(y)\,\,\text{ for }i=1,\dots, N.
			\end{align}
		The functions $U_{0}$ and $U_{i}$ are both affine and continuous, and at least one of the functions $U_{i}$ is not identically zero. Also, note that 
		\begin{align*}
			\eta(x,y,\bm a) = U_{0}(x,y) + \sum_{i=1}^{N}a_{i}U_{i}(x,y).
		\end{align*}
		Then for any fixed $\bm a\in \Re^{N}_{+}$ the mapping $(x,y)\mapsto \eta(x,y,\bm a)$ is affine and continuous, and so is the function $\bm a\mapsto \eta(x,y,\bm a)$ for any fixed $(x,y)\in X\times X$. By the minimax theorem (Theorem N$^{\prime}$ in \cite{kneser1952} or Theorem 4.2 in \cite{sion1958}) we therefore know that
		\begin{align}\label{equation: minmax maxmin eta}
			\inf_{\bm a\in \Re^{N}_{+}}\max_{(x,y)\in X\times X}\eta(x,y,\bm a) = \sup_{(x,y)\in X\times X}\inf_{\bm a\in \Re^{N}_{+}}\eta(x,y,\bm a).
		\end{align}
		Consider the function $\F\colon X\times X\to\Re\cup\{-\infty\}$ defined by 
		\[\F(x,y)=\inf_{\bm a\in\Re_{+}^{N}}\eta(x,y,\bm a).\]
		Then $\F$ is an upper semicontinuous function and, since $\F(x,x)=0$, we have that $\dom \F = \{(x,y)\in X\times X:\F(x,y)>-\infty\}\neq \emptyset$. Let $(x_{0},y_{0})\in\dom \F$. We must have that
		\begin{align*}
			\sup_{(x,y)\in X\times X}\F(x,y) = \sup\{\F(x,y):(x,y)\in X\times X, \F(x,y)\geq \F(x_{0},y_{0})\}.
		\end{align*}
		Because of upper semicontinuity of $\F$ and compactness of $X\times X$, the set $\{(x,y)\in X\times X, \F(x,y)\geq \F(x_{0},y_{0})\}$ is a nonempty compact subset of $\dom \F$. Since the restriction of $\F$ to any subset of $\dom \F$ makes the function real-valued, it follows from the Weierstrass theorem (see Theorem 2.43 in \cite{aliprantis2006})  that the maximization problem
		\begin{align*}
			\max\{\F(x,y):(x,y)\in X\times X, \F(x,y)\geq \F(x_{0},y_{0})\} 
		\end{align*}
		has a solution, and thus the supremum on the right-hand side of (\ref{equation: minmax maxmin eta}) becomes a maximum, that is, 
		\begin{align}\label{equation: minimax sup replace with max}
			\inf_{\bm a\in \Re^{N}_{+}}\max_{(x,y)\in X\times X}\eta(x,y,\bm a)= \max_{(x,y)\in X\times X}\inf_{\bm a\in \Re^{N}_{+}}\eta(x,y,\bm a).
		\end{align}
		 To establish (\ref{equation: minimax property}), it remains to show that the infimum on the left-hand side in (\ref{equation: minimax sup replace with max}) is attained.
		 
		 	In view of (\ref{equation: minimax sup replace with max}), let $(\bm a_{n})$ be a sequence in $\Re^{N}_{+}$ with 
		\begin{align*}
			\lim_{n\to \infty}\max_{(x,y)\in X\times X}\eta(x,y,\bm a_{n}) =  \inf_{\bm a\in\Re^{N}_{+}}\left (\max_{(x,y)\in X\times X}\eta(x,y,\bm a)\right),
		\end{align*}
		which is a real number. Since the convex cone generated by the functions $U_{1},\dots,U_{N}$ lies in the finite-dimensional subspace spanned by the set  $\{U_{1},\dots, U_{N}\}$, for $\bm a_{n} = (a_{1n},\dots,a_{Nn})$ we know by Carathéodory's theorem for cones (see e.g., the proof of Corollary 5.25 in \cite{aliprantis2006}, or part (i) of Proposition A.35 in \cite{schmudgen2017}) that each  $V_{n}\colon X\times X\to\Re$ given by
	 \begin{align*}
	 	V_{n}(x,y) = \sum_{i=1}^{N}a_{in}U_{i}(x,y)
	 \end{align*}
	can also be expressed, by a suitable choice of a subset  $I_{n}\subseteq\{1,\dots,N\}$ of indices, with linearly independent functions $\{U_{i}:i\in I_{n}\}$ and $ \bar \abold_{n}\in \Re^{N}_{+}$ so that
	 \begin{align}\label{equation: expression for Vn with LI U}
	 	V_{n}(x,y) = \sum_{i\in I_{n}}\bar a_{in}U_{i}(x,y).
	 \end{align}
	 Since there are at most $2^{N}-1$ choices for linearly independent subsets of $\{U_{1},\dots,U_{N}\}$, for at least one subset $I^{\ast}\subseteq \{1,\dots,N\} $ of indices there exists a subsequence of $(V_{n})$ such that each term in it is expressed as the linear combination of $\{U_{i}:i\in I^{\ast}\}$ with nonnegative coefficients. So we assume without loss of generality that the sequence $(V_{n})$ is expressed in such way as $V_{n} = \sum_{i\in I^{\ast}}\bar a_{in}U_{i}$. Note that we also have 
		\begin{align}\label{equation: limit of max eta}
			\lim_{n\to \infty}\max_{(x,y)\in X\times X}\eta(x,y,\bar\abold_{n}) =  \inf_{\bm a\in\Re^{N}_{+}}\left (\max_{(x,y)\in X\times X}\eta(x,y,\bm a)\right).
		\end{align}
If the sequence $(\bar \abold_{n})$ is unbounded, then we may assume without loss of generality that  $\norm{\bar \abold_{n}}$, taken as any norm in $\Re^{I^{\ast}}$, increases to $+\infty$. By compactness of the unit ball of $\Re^{I^{\ast}}$ in any norm topology, and by passing to a convergent subsequence if needed, to save on notation we suppose that $\lim_{n\to\infty} \norm{\bar \abold_{n}}^{-1}\bar \abold_{n}=\hat \abold $, for some $\hat  \abold \in\Re^{I^{\ast}}_{+}$ with $\norm{\hat\abold}=1$. Combining this observation with the limit in (\ref{equation: limit of max eta}) we must have that, for all $(x,y)\in X\times X$,
	\begin{align}\label{equation: inequality involving eta normalized unbdd}
		\sum_{i\in I^{\ast}}\hat a_{i}U_{i}(x,y)& =\lim_{n\to \infty}\left(\norm{\bar \abold_{n}}^{-1}U_{0}(x,y)+  \sum_{i\in I^{\ast}}\norm{\bar \abold_{n}}^{-1}\bar a_{in}U_{i}(x,y)\right)\notag\\	
 	  &\leq \lim_{n\to\infty }\norm{\bar \abold_{n}}^{-1} \max_{(x,y)\in X\times X}\eta(x,y,\bar \abold_{n})\notag \\
 	 & = \lim_{n\to\infty}\norm{\bar \abold_{n}}^{-1}\lim_{n\to\infty}\max_{(x,y)\in X\times X}\eta(x,y,\bar \abold_{n})\notag \\
 	  	&= 0.
	\end{align}
	If we interchange the roles of $x$ and $y$, and use the fact that $U_{i}(y,x)=-U_{i}(x,y)$, it follows from (\ref{equation: inequality involving eta normalized unbdd}) that  $\sum_{i\in I^{\ast}}\hat a_{i}U_{i}(x,y) = 0$. Since $\hat \abold\neq 0$ we have a contradiction with the linear independence of the set $\{U_{i}: i\in I^{\ast}\}$. Therefore, $(\bar \abold_{n})$ is a bounded sequence, so we let $(\bar \abold_{n_{k}})$ denote a convergent subsequence with limit $\abold$. Hence, since $X\times X$ is compact, and $\eta$ is a continuous function, by the maximum theorem (see Theorem 17.31 in \cite{aliprantis2006}), we now obtain that	
	\begin{align}\label{equation: maximum eta attained}
			\max_{(x,y)\in X\times X}\eta(x,y,\abold) = \lim_{k\to \infty}\max_{(x,y)\in X\times X}\eta(x,y,\bar\abold_{n_{k}}) =  \inf_{\bm a\in\Re^{N}_{+}}\left (\max_{(x,y)\in X\times X}\eta(x,y,\bm a)\right).
		\end{align}	
		This establishes the minimax property in (\ref{equation: minimax property}) of the function $\eta$.
	\end{proof}
\end{lemma}

Now turn to the proof of Theorem \ref{theorem: duality V}. Because of Lemma \ref{lemma: minimax duality}, let $\abold\in\Re^{N}_{+}$ be such that (\ref{equation: maximum eta attained}) holds. Then the function $e(x)= u_{0}(x)-\sum_{i=1}^{N}a_{i}u_{i}(x)$ has
	\begin{align}\label{equation: expression for oscillation minimax}
		\omega_{e}(X)  = \max_{(x,y)\in X\times X}\eta(x,y,\abold) = \max_{(x,y)\in X\times X} \F(x,y),
	\end{align}
	where the second equality in (\ref{equation: expression for oscillation minimax}) follows from (\ref{equation: minimax sup replace with max}). Since
	\begin{align*}
		\F(x,y) = \begin{cases}
			u_{0}(y)-u_{0}(x),&\text{ if }u_{i}(x)\geq u_{i}(y)\text{ for all }i=1,\dots,N\\
			-\infty,&\text{ if }u_{i}(x)<u_{i}(y)\text{ for some }i\in \{1,\dots,N\}		 
		\end{cases}
			\end{align*}
			we obtain the duality formula (\ref{equation: duality V}).


\subsection{Proof of Theorem \ref{theorem: aggregation finite semistrong compact}}

To show that (i) implies (ii), since the assumptions of Theorem \ref{theorem: duality V} are met, we know that there exists $\abold\in \Re^{N}_{+}$ with the property that the function $e\colon X\to \Re $ as defined implicitly in the approximate aggregation formula $u_{0}(x)=\sum_{i=1}^{N}a_{i}u_{i}(x)+e(x)$ has its oscillation given by $\omega_{e}(X) = \max\{u_{0}(y)-u_{0}(x):(x,y)\in\V \} $. By $\epsilon$-Semistrong Pareto we have that $u_{0}(y)-u_{0}(x)\leq \epsilon$ for all $(x,y)\in \V$, thus establishing (ii).
	
	To prove that (ii) implies (i),  note  that if $u_{i}(x) \geq  u_{i}(y) $ for all $i=1,\dots,N$, then $w(x)\geq w(y)$, and thus
\begin{align*}
	u_{0}(x) - u_{0}(y) &\geq u_{0}(x)  - w(x) - (u_{0}(y) - w(y))\geq -\frac{\epsilon}{2} -\frac{\epsilon}{2}, 
\end{align*}
so that $u_{0}(x)\geq u_{0}(y)-\epsilon$.


\subsection{Proof of Theorem \ref{theorem: semistrong without compactness}}


The proof that (ii) implies (i) follows the same reasoning as in Theorem \ref{theorem: aggregation finite semistrong compact}. Therefore, we only need to prove that (i) implies (ii). Here we assume that at least one of the functions $u_{i}$ is not constant, for otherwise the choice of $a_{i}=0$ for $i=1,\dots,N$ is possible, and $\epsilon$-Semistrong Pareto gives $\omega_{e}(X)=\omega_{u_{0}}(X)\leq \epsilon$.
		
		We first define $W = X\times X$, and note that if $\{V_{1},\dots,V_{M}\}$ is a set of linearly independent functions from $W$ to $\Re$ then there exists a subset $W_{0}$ of $W$ with $M$ elements  such that
		\begin{align}\label{equation: linear independence of Vs}
			 \{(V_{1}(\w),\dots,V_{M}(\w)):\w\in W_{0}\}\quad \text{is a linearly independent subset of }\Re^{M}.
		\end{align}
		The choice of $W_{0}$ is possible because, using the linear independence of the set $\{V_{1},\dots,V_{M}\}$,  the orthogonal complement of the vector space spanned by $\{(V_{1}(\w),\dots,V_{M}(\w)):\w\in W\}$ is the trivial subspace $\{0\}$ of $\Re^{M}$.
		
		Now define the $N+1$ functions $U_{0},U_{i}\colon W\to\Re$, $i=1,\dots,N$, as in (\ref{equation: functions U0 Ui defined}). For each finite subset $D$ of $W$, there corresponds a finite set $X_{D}\subseteq X$  with the property that $D \subseteq X_{D}\times X_{D}\subseteq \co(X_{D}\times X_{D})=\co X_{D}\times \co X_{D}$.\footnote{We use the standard notation $\co A$ to represent the convex hull of a subset $A$ of a real vector space.} Here, we take $X_{D}$ to be the set of all elements of $X$ that appear in any pair in $D$. Define $W_{D} = X_{D}\times X_{D}$. Hence, for each finite set $D\subseteq W$, the restrictions of $U_{0}$ and $U_{i}$ to $\co W_{D}$, namely, $\restr{U_{0}}{\co W_{D}},\restr{U_{i}}{\co W_{D}}\colon \co W_{D}\to \Re $, are induced by the restrictions of the functions $u_{0}$ and $u_{i}$ to $\co X_{D}$. Owing to the mixture-preserving property, the functions $\restr{u_{0}}{\co X_{D}},\restr{u_{i}}{\co X_{D}}\colon \co X_{D}\to \Re $ are continuous with respect to the standard Euclidean topology on $\co X_{D}$ when this set is viewed as a finite-dimensional convex polytope. This topology is the only Hausdorff linear topology on the vector space spanned by $X_{D}$ (e.g., Theorem 5.21 in \cite{aliprantis2006}). Moreover, as the convex hull of finitely many points in $X$, the set $\co X_{D}$ is compact in that same topology (see Corollary 5.30 in \cite{aliprantis2006}). Therefore, as the $\epsilon$-Semistrong Pareto axiom also holds in the restricted subset $\co X_{D}$ of $X$, Theorem \ref{theorem: aggregation finite semistrong compact} applies and we can find $\abold_{D} = (a_{1D},\dots,a_{ND})\in \Re^{N}_{+}$ and a function $e_{D}\colon \co X_{D}\to \Re$, with $\omega_{e_{D}}(\co X_{D})\leq \epsilon$, such that
		\begin{align}\label{equation: representation difference fixed D}
			U_{0}(x,y) + \sum_{i=1}^{N}a_{iD}U_{i}(x,y) =  \eta(x,y,\abold_{D}) \quad\text{for all }(x,y)\in \co W_{D},
		\end{align}
		where $\eta $ is defined as in (\ref{equation: eta defined}), and we have $e_{D}(y) - e_{D}(x) =  \eta(x,y,\abold_{D})$. For notation, when $\w=(x,y)$ we write $\eta_{D}(\w)$ instead of $ \eta(x,y,\abold_{D})$, and note that $\sup_{\w\in \co W_{D}}\abs{\eta_{D}(\w)}\leq \epsilon$.
		
		The set of functions $\{U_{1},\dots,U_{N}\}$, when viewed as a subset of the vector space $\Re^{W}$, has at most $2^{N}-1$ linearly independent subsets. Let $\mathcal I$ be the family of all sets $I$ of indices $i\in \{1,\dots,N\}$ for which the set $\{U_{i}: i\in I \}$ is linearly independent in $\Re^{W}$. The family $\mathcal I$ is nonempty since at least one of the $u_{i}$'s is not constant. For each $I\in\mathcal I$ and $\w\in W$, the function that maps $i\in I$ to  $U_{i}(\w)$ is denoted by $\bm U(\w)$, and viewed as a vector in $\Re^{\abs{I}}$. We also define the square matrix of order $\abs{I}$
			\begin{align}\label{equation: alternant matrix}
			\bm M_{I} = \begin{bmatrix}
			\llongdash &\bm U(\w_{1})&\rlongdash\\
			\llongdash  &	\bm U(\w_{2})&\rlongdash\\
				&\vdots &\\
			\llongdash &	 \bm U(\w_{\abs{I}})&\rlongdash
			\end{bmatrix},
		\end{align}
		where the set $W_{I} = \{\w_{1},\dots,\w_{\abs{I}}\}$ is chosen as mentioned in the argument above justifying the condition in (\ref{equation: linear independence of Vs}). By our choice of $W_{I}$, the alternant matrix $\bm M_{I}$ is nonsingular. We also define the vectors
		\begin{align*}
			\bm U_{0I} = \begin{bmatrix}
				U_{0}(\w_{1})\\
				U_{0}(\w_{2})\\
				\vdots\\
				U_{0}(\w_{\abs{I}})
			\end{bmatrix}\quad\text{and}\quad \bm \eta_{D}= \begin{bmatrix}
					\eta_{D}(\w_{1})\\
					\eta_{D}(\w_{2})\\
					\vdots\\
					\eta_{D}(\w_{\abs{I}})
				\end{bmatrix},
		\end{align*}
		both in $\Re^{\abs{I}}$.

	Let $D_{0}$ denote the set $\bigcup_{I\in\mathcal I}W_{I}$. The family of all finite subsets $D$ of $W$ with $D_{0}\subseteq D$ is denoted by $\mathcal D$. Note that the set $\mathcal D$ is directed by the superset relation $\supseteq$.\footnote{Recall that a direction on $\mathcal D$ is a reflexive and transitive binary relation on $\mathcal D$ for which any pair of elements of $\mathcal D$ has an upper bound in $\mathcal D$ (see \cite{aliprantis2006}, p.\ 29). Here, the union of two elements in $\mathcal D$ is also a member of $\mathcal D$, and thus becomes the relevant upper bound with the partial order $\supseteq$.} For any fixed element $D\in\mathcal D$, since $D_{0}\subseteq D$ the subsets of indices corresponding to linearly independent subsets of $\left \{\restr{U_{1}}{\co W_{D}},\dots,\restr{U_{N}}{\co W_{D}} \right\}$ are precisely the elements of $\mathcal I$. Recall that a finitely generated cone in a vector space is the union of the cones generated by the linearly independent subsets of the generating set.\footnote{This assertion is a consequence of Carathéodory's theorem for cones. See, for instance, the proof of Corollary 5.25 in \cite{aliprantis2006}, or item (i) of Proposition A.35 in \cite{schmudgen2017}. These were also mentioned in the proof of Theorem \ref{theorem: duality V}.} Hence, viewing $\left \{\restr{U_{1}}{\co W_{D}},\dots,\restr{U_{N}}{\co W_{D}} \right\}$ as the generating set we can replace the vector $\abold_{D}$ in (\ref{equation: representation difference fixed D}) by $\bar \abold_{D}$ that is supported on a set $I\in \mathcal I$ of indices. Equation (\ref{equation: representation difference fixed D}) remains valid with $\abold_{D}$ replaced with $\bar \abold_{D}$ since $ \sum_{i=1}^{N}a_{iD}U_{i} =  \sum_{i=1}^{N}\bar a_{iD}U_{i}$.  It follows from this observation that
			\begin{align*}
				\restr{U_{0}}{\co W_{D}} + \sum_{i=1}^{N}\bar a_{iD}\restr{U_{i}}{\co W_{D}} = \eta_{D}.
			\end{align*}
		If we identify $\bar \abold_{D}$ with a vector in $\Re^{\abs{I}}$ then we can solve for $\bar \abold_{D}$ with the matrix $\bm M_{I}$ and the vector $\bm U_{0I}$ using the expression $\bm M_{I}^{-1}\bm \eta_{D} - \bm M_{I}^{-1}\bm U_{0I} $. Since $\norm{\bm \eta_{D}}_{\infty}\leq \epsilon$ we obtain the estimate
		\begin{align}\label{equation: inequality estimate bound for aD}
			\norm{\bar \abold_{D}}_{\infty}\leq \max\left \{\norm{\bm M_{I}^{-1}\bm U_{0I}}_{\infty} + \epsilon\norm{|\bm {M}_{I}^{-1}|\indic_{I}}_{\infty}:I\in \mathcal I \right \},
		\end{align}
		where $|\bm{M}_{I}^{-1}|$ denotes the matrix of entrywise absolute values and the function $\indic_{I}$ with domain $I$ is viewed as a vector whose dimension is compatible with that of $\bm M_{I}^{-1}$. The maximum in (\ref{equation: inequality estimate bound for aD}) is attained because $\mathcal I$ is finite. Therefore, the net	$(\bar \abold_{D})_{D\in\mathcal D}$ in $\Re_{+}^{N}$ is bounded. 
		
		By the characterization of compact sets of topological spaces (Theorem 2.31 in \cite{aliprantis2006}), it follows that $(\bar \abold_{D})_{D\in\mathcal D}$ admits a convergent subnet. Hence for some set $\Gamma$ directed by a reflexive and transitive binary relation which we denote by $\unrhd$, and some function $\varphi\colon\Gamma \to \mathcal D$ with the property that for every $D_{1}\in \mathcal D$ there exists $\gamma_{1}\in \Gamma$ such that $\gamma\unrhd \gamma_{1}$ implies $\varphi(\gamma)\supseteq D_{1}$, the net $(\hat\abold_{\gamma})$ defined so that $\hat \abold_{\gamma} = \bar \abold_{\varphi(\gamma)}$ converges.
		
		Put $\abold = \lim_{\gamma}\hat \abold_{\gamma} \in\Re^{N}$. Fix any $(x,y)\in W$, and let $D_{1}\in\mathcal D$ be a set containing the pair $(x,y)$. Then for some $\gamma_{1}\in \Gamma$ we know that $\gamma\unrhd \gamma_{1}$ implies  $\varphi(\gamma)\supseteq D_{1}$.  Note also that the tail  $(\hat \abold_{\gamma})_{\gamma\unrhd \gamma_{1}} $, viewed as a further subnet, also converges to the \textit{same} limit $\abold$. In particular, we also have that
		\begin{align*}
		 	\lim_{\gamma} \left [U_{0}(x,y) -\eta_{\varphi(\gamma)}(x,y)\right] =-\lim_{\gamma} \sum_{i=1}^{N}\hat a_{i\varphi(\gamma)}U_{i}(x,y) =- \sum_{i=1}^{N}a_{i}U_{i}(x,y),
		 \end{align*}
	 and thus the net $(\eta_{\varphi(\gamma)}(x,y))_{\gamma\unrhd \gamma_{1}}$ converges to some $\eta(x,y)\in\Re$, which has $\abs{\eta(x,y)}\leq \epsilon$ as every term of that net is bounded in absolute value by the same $\epsilon$. Conclusion: $u_{0}(y) - u_{0}(x)+ \sum_{i=1}^{N}a_{i}[u_{i}(x)-u_{i}(y)]=\eta(x,y)$. Upon defining $e(x) = u_{0}(x)-\sum_{i=1}^{N}a_{i}u_{i}(x)$ we establish the assertion in (ii).


\subsection{Proof of Theorem \ref{theorem: aggregation finite indifference}}


The statements in (i) and (ii) can be written with $\epsilon$-Semistrong Pareto and the aggregation with nonnegative coefficients when we replace $\U$ with the set $\{u\in \Re^{X}:u\in \U\text{ or }-u\in \U\}$. The fact that both statements are equivalent now follows from Theorem \ref{theorem: semistrong without compactness} and simple algebra.


\subsection{Proof of Theorem \ref{theorem: aggregation with sequential epsilon strong}}


Let the functions $U_{0}$ and $U_{i}$ be defined as in (\ref{equation: functions U0 Ui defined}). We first show that (ii) implies (i). From the representation, we know that
		\begin{align*}
			U_{0}(x,y)+\sum_{i=1}^{N}a_{i}U_{i}(x,y)\leq \epsilon\quad\text{for all }x,y\in X,
		\end{align*}
		so in particular the same inequality obtains with sequences $(x_{n})$ and $(y_{n})$ in $X$. Now assume that there exists $i\in \{1,\dots,N\}$ such that $U_{i}(x_{n},y_{n})>0$ for all $n$, which means that $u_{i}(x_{n})-u_{i}(y_{n})>0$. Hence
		\begin{align}\label{equation: liminf ineq necessity}
		 \frac{u_{0}(y_{n})-u_{0}(x_{n})-\epsilon }{u_{i}(x_{n})-u_{i}(y_{n})} + 	\sum_{\substack{j=1\\ j\neq i}}^{N}a_{j}\left(\frac{u_{j}(x_{n})-u_{j}(y_{n})}{u_{i}(x_{n})-u_{i}(y_{n})}\right )\leq -a_{i}.
		\end{align}
		Because the Pareto weights are positive, the right-hand side of (\ref{equation: liminf ineq necessity}) is negative, and so is its limit inferior. If
		\begin{align*}
		 \liminf_{n\to\infty}\frac{u_{0}(y_{n})-u_{0}(x_{n})-\epsilon}{u_{i}(x_{n})-u_{i}(y_{n})}\geq 0\quad\text{and}\quad	\liminf_{n\to\infty}\frac{u_{j}(x_{n})-u_{j}(y_{n})}{u_{i}(x_{n})-u_{i}(y_{n})}\geq 0,
		\end{align*}
		then, combining the superadditivity of the limit inferior operator with the property about the sign of each $a_{j}$ and the expression (\ref{equation: liminf ineq necessity}), we obtain that
		\begin{align}\label{equation: strict ineq liminf proof}
			0&\leq \liminf_{n\to\infty}\frac{u_{0}(y_{n})-u_{0}(x_{n})-\epsilon }{u_{i}(x_{n})-u_{i}(y_{n})}  + \sum_{\substack{j=1\\ j\neq i}}^{N}a_{j}\left(\liminf_{n\to\infty}\frac{u_{j}(x_{n})-u_{j}(y_{n})}{u_{i}(x_{n})-u_{i}(y_{n})}\right) \notag\\
					&\leq \liminf_{n\to\infty}\Bigg(\frac{u_{0}(y_{n})-u_{0}(x_{n})-\epsilon }{u_{i}(x_{n})-u_{i}(y_{n})} + \sum_{\substack{j=1\\ j\neq i}}^{N}a_{j}\left(\frac{u_{j}(x_{n})-u_{j}(y_{n})}{u_{i}(x_{n})-u_{i}(y_{n})}\right) \Bigg)\notag\\
					&<0.
		\end{align}
		This is a contradiction, where the strict inequality in (\ref{equation: strict ineq liminf proof}) is a consequence of the remark following (\ref{equation: liminf ineq necessity}). This shows part (b) of the Sequential $\epsilon$-Strong Pareto condition. Part (a) of the same axiom is a consequence of the sign of the Pareto weights as in Theorem \ref{theorem: aggregation finite semistrong compact}.
		
		To show that (i) implies (ii) we first show that for any fixed $i\in \{1,\dots,N\}$, there exist a number $a_{0}\geq 0$ and a list $\abold_{-i}$ of $N-1$ nonnegative numbers  $a_{j}$, $j\in\{1,\dots,N\}\setminus\{i\}$, such that 
		\begin{align*}
			a_{0}U_{0}(x,y) + \sum_{\substack{j=1\\ j\neq i}}^{N}a_{j}U_{j}(x,y) + U_{i}(x,y)\leq a_{0}\epsilon\quad\text{for all }x,y\in X.
		\end{align*}
		To this end we consider, for each $n\in \Na$, the function $\theta_{n}\colon W\times \left [\frac{1}{n},1\right]\times [0,1]^{N}\to \Re $ which maps each $(x,y)\in W=X\times X$ and $\bm a = (a_{i},a_{0},\bm a_{-i})\in  \left[\frac{1}{n},1\right]\times [0,1]\times [0,1]^{N-1}$  to 
		\begin{align*}
			\theta_{n}(x,y,\bm a) = a_{0}[U_{0}(x,y)-\epsilon] +  \sum_{\substack{j=1\\ j\neq i}}^{N}a_{j}U_{j}(x,y)  + a_{i}U_{i}(x,y).
		\end{align*}
		For any fixed $(x,y)$ the mapping $\bm a\mapsto \theta_{n}(x,y,\bm a)$ is affine and continuous, and for any given $\bm a$ the mapping $(x,y)\mapsto \theta_{n}(x,y,\bm a)$ is affine. Since the convex set $A_{n} = \left[\frac{1}{n},1\right]\times [0,1]^{N}$ is compact when endowed with the product topology and the real line is equipped with its natural topology, and the set $W$ is also convex, we can apply the minimax theorem (Theorem N$^{\prime}$ in \cite{kneser1952} or Theorem 4.2 in \cite{sion1958}) to obtain that 
		\begin{align}\label{equation: minimax sequential}
			\sup_{(x,y)\in W}\min_{\bm a\in A_{n}}\theta_{n}(x,y,\bm a)= \inf_{\bm a\in A_{n}}\sup_{(x,y)\in W}\theta_{n}(x,y,\bm a).
		\end{align}
		
		Since $A_{n}\subseteq A_{n+1}$, if the common value in (\ref{equation: minimax sequential}) is negative or zero for some $\bar n$, then it is also negative or zero for all $n\geq \bar n$. We now prove that such a threshold exists. To achieve this it suffices to show that we cannot find a strictly increasing sequence $(n_{k})$ of natural numbers with the property that $\sup_{(x,y)\in W}\min_{\bm a\in A_{n_{k}}}\theta_{n_{k}}(\bm a,x,y)>0$ for all $k$. If such a sequence in $\Na$ exists then we can find sequences $(x_{k})$ and $(y_{k})$ in $X$ with 
		\begin{align*}
			\min_{\bm a\in A_{n_{k}}}\Bigg  \{a_{0}[u_{0}(y_{k}) - u_{0}(x_{k})-\epsilon] +  \sum_{\substack{j=1\\ j\neq i}}^{N}a_{j}[u_{j}(x_{k})- u_{j}(y_{k})] + a_{i}[u_{i}(x_{k})- u_{i}(y_{k})]\Bigg\}>0.
		\end{align*}
		Setting $a_{0} = 0$ and $a_{j}=0$ for $j\neq i,0$, we have that $u_{i}(x_{k})>u_{i}(y_{k})$ for all $k$. For $a_{i} = \frac{1}{n_{k}}$, $a_{j}=1$ for some $j$ (but $=0$ for the remaining indices),  and $a_{0} = 0$ we deduce that
		\begin{align*}
			\frac{u_{j}(x_{k})- u_{j}(y_{k})}{u_{i}(x_{k})- u_{i}(y_{k})}>-\frac{1}{n_{k}} \text{ for all }k,  
					\end{align*}
		and thus
		\begin{align*}
			 \liminf_{k\to\infty}\frac{u_{j}(x_{k})- u_{j}(y_{k})}{u_{i}(x_{k})- u_{i}(y_{k})}\geq 0,
		\end{align*}			
		since $\frac{1}{n_{k}}\to 0$. Using a similar argument, we also obtain that 
		\begin{align*}
			\liminf_{k\to\infty}\frac{u_{0}(y_{k})- u_{0}(x_{k})-\epsilon}{u_{i}(x_{k})- u_{i}(y_{k})}\geq 0.
		\end{align*}
		This is a contradiction with the Sequential $\epsilon$-Strong Pareto axiom. 
		
		We now know that there exists a natural number $n$ such that, upon defining the lower semicontinuous function $\G\colon A_{n}\to \Re\cup \{+\infty\}$ by
		\begin{align*}
			\G(\bm a)=\sup_{x,y\in X}\Bigg\{a_{0}[u_{0}(y) - u_{0}(x)-\epsilon] + \sum_{\substack{j=1\\ j\neq i}}^{N}a_{j}[u_{j}(x)- u_{j}(y)] + a_{i}[u_{i}(x)- u_{i}(y)]\Bigg\},
		\end{align*}
		 we have that
		\begin{align}\label{equation: inequality with G}
			\inf_{\bm a \in A_{n}}\G(\bm a)\leq 0.
		\end{align}
		Lower semicontinuity of $\G$ follows from the fact that it is the pointwise supremum of continuous mappings. Note that in view of (\ref{equation: inequality with G}) we also know that $\dom \G  = \{\bm a\in A_{n}:\G(\bm a)<+\infty\}\neq \emptyset$. For any fixed $\abold_{0} \in\dom \G$, we must have that
		\begin{align}\label{equation: inf G attained}
			\inf_{\abold \in A_{n}}\G(\abold)= \inf\{\G(\abold): \abold \in A_{n}, \G(\abold)\leq \G(\abold_{0})\}.
		\end{align}
		Since $A_{n}$ is compact and the lower contour sets of $\G$ are closed, the set $\{ \abold \in A_{n}: \G(\abold)\leq \G(\abold_{0})\}$ is compact. Hence, by the Weierstrass theorem (see Theorem 2.43 in \cite{aliprantis2006}) applied to the restriction of $\G$ to this compact set, we know that the infimum on the right-hand side of (\ref{equation: inf G attained}) is attained at some $\bm a\in A_{n}$. Since $a_{i}>0$ we can normalize $a_{i}=1$, so there exist $a_{0}\geq 0$ and $a_{j}\geq 0$, $j\neq i,0$, with 
		\begin{align*}
			a_{0}[u_{0}(y) - u_{0}(x)-\epsilon] + \sum_{\substack{j=1\\ j\neq i}}^{N}a_{j}[u_{j}(x)- u_{j}(y)] + u_{i}(x)- u_{i}(y)\leq 0
		\end{align*}
		for all $x,y\in X$. By interchanging the roles of $x$ and $y$, we obtain
		\begin{align*}
			-a_{0}\epsilon\leq a_{0}[u_{0}(y) - u_{0}(x)]+ \sum_{\substack{j=1\\ j\neq i}}^{N}a_{j}[u_{j}(x)- u_{j}(y)] + u_{i}(x)- u_{i}(y)\leq a_{0}\epsilon.
		\end{align*}
		
		Combining the result above and Theorem \ref{theorem: semistrong without compactness}, we know that there exist lists of nonnegative numbers $(a_{0}^{0},a_{0}^{1},\dots,a_{0}^{N})$ and $(a_{i}^{0},a_{i}^{1},\dots,a_{i}^{N})$ for $i=1,\dots,N$ with $a_{0}^{0}=1$ and $a_{i}^{i} = 1$ for $i=1,\dots,N$, and such that for all $x,y\in X$
		\begin{align*}
			-a_{0}^{l}\epsilon\leq  a_{0}^{l}[u_{0}(y) - u_{0}(x)]+ \sum_{i=1}^{N}a_{i}^{l}[u_{i}(x)- u_{i}(y)]\leq a_{0}^{l}\epsilon\quad\text{for }l=0,1,\dots,N.
		\end{align*}
		Then
		\begin{align*}
			\abs{\sum_{i=1}^{N}a_{i}[u_{i}(x)- u_{i}(y)] - a_{0}[u_{0}(x) - u_{0}(y)]}\leq a_{0}\epsilon,
		\end{align*}
		where $a_{i} = \sum_{l=0}^{N}a_{i}^{l}>0$ and $a_{0} = \sum_{l=0}^{N}a_{0}^{l}>0$. The result is now a consequence of the normalization $a_{0}=1$.


\subsection{Proof of Theorem \ref{theorem: aggregation strong polyhedral}}

The proof that (ii) implies (i) is the particular case of checking the Sequential $\epsilon$-Strong Pareto axiom when the sequences involved are all constant sequences, so it follows from Theorem \ref{theorem: aggregation with sequential epsilon strong}. We only show that (i) implies (ii). Here, as a consequence of Theorem \ref{theorem: aggregation with sequential epsilon strong}, it suffices to prove that the pair $(u_{0},\U)$ satisfies Sequential $\epsilon$-Strong Pareto. 
		
	Let $\hat Z$ denote the set of all vectors $(-u_{0}(x),u_{1}(x),\dots,u_{N}(x))\in \Re^{N+1}$ where $x\in X$. It follows that the set $\hat Z$ is the convex hull of finitely many points. The set $\hat Z$ is thus a polyhedral set, that is, $\hat Z$ is the set of all solutions to a finite system of linear inequalities (see e.g., \citet[Theorem 13.16]{soltan2019}). Define $\bm  \epsilon = (\bm \epsilon(0),\bm \epsilon(1),\dots,\bm \epsilon(N))$ as the vector in $\Re^{N+1}$ with $\bm\epsilon(0)=\epsilon$ and $\bm \epsilon(i)=0$ for $i=1,\dots,N$. It follows from Theorems 13.19 and 13.20 in \cite{soltan2019} that the set
	\begin{align}\label{equation: define z epsilon}
		Z = \co( \{\hat z - \tilde z\in \Re^{N+1}: \hat z,\tilde z\in \hat Z \}\cup\{\bm \epsilon\})
	\end{align}
	is also a polyhedral set.
		
	Now we show that there are no sequences $(x_{n})$ and $(y_{n})$ in $X$ for which, for some $i\in \{1,\dots,N\}$, we have that $u_{i}(x_{n})>u_{i}(y_{n})$, $\liminf_{n\to\infty}\frac{u_{j}(x_{n})-u_{j}(y_{n})}{u_{i}(x_{n})-u_{i}(y_{n})}\geq 0$ for all $j\neq i,0$ and $ \liminf_{n\to\infty}\frac{u_{0}(y_{n})-u_{0}(x_{n})-\epsilon}{u_{i}(x_{n})-u_{i}(y_{n})}\geq 0$. Suppose, by way of contradiction, that such sequences exist. They induce a sequence $(z_{n})$ in the set $Z$ with the property that $z_{n}(i)>0$ and $\liminf_{n\to\infty}\frac{z_{n}(j)-\bm\epsilon(j)}{z_{n}(i)-\bm\epsilon(i)}\geq 0$ for all $j\in \{0,1,\dots,N\}$. Define the sets 
	\begin{align}
		J_{1} &= \left\{ j\in \{0,1,\dots, N\}: \liminf_{n\to\infty}\frac{z_{n}(j)-\bm\epsilon(j)}{z_{n}(i)-\bm\epsilon(i)} = + \infty\right\} \label{equation: define J1}\\
		J_{2}&=\left\{ j\in \{0,1,\dots, N\}: \liminf_{n\to\infty}\frac{z_{n}(j)-\bm\epsilon(j)}{z_{n}(i)-\bm\epsilon(i)} < + \infty\right\}.\label{equation: define J2}
	\end{align}
 For a fixed $j\in J_{2}\setminus\{i\}$, by extracting a convergent subsequence of quotients $\frac{z_{n}(j)-\bm\epsilon(j)}{z_{n}(i)-\bm\epsilon(i)}$  we may assume without loss of generality that the limit of $\frac{z_{n}(j)-\bm\epsilon(j)}{z_{n}(i)-\bm\epsilon(i)}$ as $n\to\infty$ exists.  This procedure may change the set $J_{2}$, thus enlarging the set $J_{1}$. In particular no $j\in J_{1}$ leaves $J_{1}$. Therefore, by repeating this procedure if needed, we can assume without loss of generality that the sequences $(x_{n})$ and $(y_{n})$ were initially given so that $\lim_{n\to\infty}\frac{z_{n}(j)-\bm\epsilon(j)}{z_{n}(i)-\bm\epsilon(i)}$ exists for all $j\in J_{2}$, and $0<\frac{z_{n}(j)-\bm\epsilon(j)}{z_{n}(i)-\bm\epsilon(i)}$ and $\lim_{n\to\infty}\frac{z_{n}(j)-\bm\epsilon(j)}{z_{n}(i)-\bm\epsilon(i)} = +\infty$ for all $j\in J_{1}$. Hence we replace the definitions in (\ref{equation: define J1}) and (\ref{equation: define J2}) with
 \begin{align}
 	J_{1} &=\left\{ j\in \{0,1,\dots, N\}:\lim_{n\to\infty}\frac{z_{n}(j)-\bm\epsilon(j)}{z_{n}(i)-\bm\epsilon(i)} = + \infty \right\}\label{equation: define new J1} \\
 	J_{2}& = \left\{ j\in \{0, 1,\dots, N\}: \lim_{n\to\infty}\frac{z_{n}(j)-\bm\epsilon(j)}{z_{n}(i)-\bm\epsilon(i)} \text{ exists}\right\}.\label{equation: define new J2}
 \end{align}
  
 First assume that $J_{1}\neq \emptyset$. Define
 \begin{align}\label{equation: define tn initial}
 	t_{n} =\frac{1}{\max\{z_{n}(j)-\bm\epsilon(j):j\in J_{1}\}}> 0.
 \end{align}  
 For any $j_{1}\in J_{1}$ we have that
	\begin{align*}
		0< t_{n}[z_{n}(j_{1})-\bm\epsilon(j_{1})] =\frac{z_{n}(j_{1})-\bm\epsilon(j_{1})}{\max\{z_{n}(j)-\bm\epsilon(j):j\in J_{1}\}}\leq 1.
	\end{align*}
  And for $j_{2}\in J_{2}$ we obtain that 
  \begin{align}\label{equation: condition limit of the j2}
  	 t_{n}[z_{n}(j_{2})-\bm\epsilon(j_{2})] = \frac{z_{n}(j_{2})-\bm\epsilon(j_{2})}{z_{n}(i)-\bm\epsilon(i)}\frac{z_{n}(i)-\bm\epsilon(i)}{\max\{z_{n}(j)-\bm\epsilon(j):j\in J_{1}\}}\to 0
  \end{align}
  as $n\to\infty$, since $\lim_{n}t_{n}[z_{n}(i)-\bm\epsilon(i)]=0$ owing to (\ref{equation: define new J1}), and for $j_{2}\in J_{2}$ the term $ \frac{z_{n}(j_{2})-\bm\epsilon(j_{2})}{z_{n}(i)-\bm\epsilon(i)}$ is bounded. The mapping $j_{1}\mapsto  t_{n}[z_{n}(j_{1})-\bm\epsilon(j_{1})]$ induces a bounded sequence in $\Re^{\abs{J_{1}}}$. By passing to a subsequence if needed, we may assume that the sequence itself converges. Note that one coordinate in each term of such sequence in $\Re^{\abs{J_{1}}}$ equals $1$ infinitely often, and thus the sequence in $\Re^{N+1}$ defined by the mapping that associates each $j=0,1,\dots,N$ with $  t_{n}[z_{n}(j)-\bm\epsilon(j)]$ converges to a nonzero vector in $\Re^{N+1}$ whose components are nonnegative. 
  
  When $J_{1}= \emptyset$, we replace the definition of $t_{n}$ in  (\ref{equation: define tn initial}) with
 \begin{align*}
 	t_{n}  =\frac{1}{z_{n}(i)-\bm\epsilon(i)}> 0.
 \end{align*}
Here, a conclusion similar to that in the previous paragraph holds. Because of (\ref{equation: define new J2}), this follows from the fact that the components of $t_{n}[z_{n}-\bm\epsilon]$ converge to nonnegative numbers,  and the component corresponding to $i$ equals $1$ along the sequence. Consequently, $\lim_{n\to\infty}t_{n}[z_{n}-\bm \epsilon]$ is a nonzero vector with nonnegative coordinates.    
  
Therefore, whenever there is a violation of Sequential $\epsilon$-Strong Pareto, we can find a nonzero vector in $\cl \cone (Z-\bm \epsilon)$ with nonnegative components. Now note that the set  $ Z-\bm \epsilon$ is a convex set and contains the origin.  It is also a polyhedral set since it is the difference between two polyhedral sets (Theorem 13.20 in \cite{soltan2019}). Because polyhedral sets have a dual expression as the sum of the convex hull of finitely many points and a finitely generated cone (see Theorem 13.16 in \cite{soltan2019}), we must have that $\cone( Z-\bm \epsilon) $ is also a polyhedral set, and thus closed. As a result, $\cl \cone( Z-\bm \epsilon)  = \cone( Z-\bm \epsilon)$. Since for some $x,y\in X$ the coordinates of the nonzero vectors in $\cone (Z-\bm\epsilon)$ are a scalar multiple of the differences of the form $u_{0}(y)- u_{0}(x)-\epsilon $ or $u_{i}(x)-u_{i}(y)$, we conclude that if there is a violation of the Sequential $\epsilon$-Strong Pareto condition, then there is also a violation of $\epsilon$-Strong Pareto.

\subsection{Proof of Proposition \ref{proposition: SEU epsilon}}


We first show that (i) implies (ii). By Example \ref{example SEU 2}, $\epsilon_{1}$-Lottery Pareto is equivalent to $\epsilon_{1}$-Semistrong Pareto when restricted to the setting of vN-M utilities with lotteries on $C$ with finite support, where the utility functions on prizes are $v_{j}$, $j=0,1,\dots,N$. Hence, it follows from Theorem \ref{theorem: semistrong without compactness}, when applied to a subdomain of the set of Savage acts, that for some $a_{1},\dots,a_{N}\geq 0$ and $b$,
		\[
		 \sup\Bigg\{\left | v_{0}(c)-\sum_{i=1}^{N}a_{i}v_{i}(c)-b\right |:c\in C\Bigg\}\leq \frac{\epsilon_{1}}{2}.
		\]
		 This is the statement about $w_{0}$ in (ii). 
		 
		 With the notation of Example \ref{example SEU 2}, let $\P = \{\bm P(E)\in \Re^{N+1}:E\in \Sigma_{S}\}$. The set $\mathcal P$ is the set of all probability profiles, including the probabilistic assessments of the events by the DM and the individuals. By Lyapunov's theorem (Theorem 13.33 in \cite{aliprantis2006}), we know that $\mathcal P$ is a compact convex subset of $\Re^{N+1}$.  We show that if $\bm t = (t_{0},t_{1},\dots,t_{N})\in \mathcal P$ then 
		 \begin{align}\label{equation: inequality min epsilon2}
		 	t_{0}\geq \min_{1\leq i\leq N} t_{i}-\frac{\epsilon_{2}}{2}.
		 \end{align}
		 Suppose by way of contradiction that for some $\bm t\in \mathcal P$ and $i^{\ast}\in \{1,\dots,N\}$ we have $t_{0}+\frac{\epsilon_{2}}{2} <t_{i^{\ast}}\leq t_{i}$ for all $i=1,\dots,N$. Now let $E\in \Sigma_{S}$ be such that $\bm t = \bm P(E)$. Using Lyapunov's theorem we can find $F\in \Sigma_{u}$ such that $P_{j}(F)=t_{i^{\ast}}$ for all $j=0,1,\dots,N$. In particular, we have that $P_{i}(E)\geq P_{i}(F)$ and thus $u_{i}(c^{\ast}Ec_{\ast})\geq u_{i}(c^{\ast}Fc_{\ast})$ for all $i=1,\dots,N$. Moreover,
		 \begin{align}\label{equation: inequality strict epsilon2}
		 	P_{0}(E)+\frac{\epsilon_{2}}{2}<P_{0}(F)
		 \end{align}
		 so that $\frac{\epsilon_{2}}{2}<P_{0}(F)$. Hence by $\frac{\epsilon_{2}}{2}$-Likelihood Dominance for $F_{\tfrac{\epsilon_{2}}{2}}\subseteq F$ we must have $P_{0}(E)\geq P_{0}\left(F\setminus F_{\tfrac{\epsilon_{2}}{2}}\right)$, that is, $u_{0}(c^{\ast}Ec_{\ast})\geq u_{0}(c^{\ast}Gc_{\ast})$ when $G = F\setminus F_{\tfrac{\epsilon_{2}}{2}}$. This contradicts the inequality in (\ref{equation: inequality strict epsilon2}).
		
		Now denote by $\Delta(N)$ the set of all $\lambda\in\Re^{N}$ with $\lambda_{i}\geq 0$ and $\sum_{i=1}^{N}\lambda_{i}=1$. If
		\begin{align*}
			\max_{\bm t \in \mathcal P}\min_{\lambda\in\Delta(N)}\Bigg\{-t_{0}  +\sum_{i=1}^{N}\lambda_{i}t_{i}\Bigg\}>\frac{\epsilon_{2}}{2}
					\end{align*}
		then for some $ \bm t\in \mathcal P$ we have that $t_{0}+\frac{\epsilon_{2}}{2}<\sum_{i=1}^{N}\lambda_{i}t_{i}$ for all $\lambda\in\Delta(N)$, so in particular $t_{0}+\frac{\epsilon_{2}}{2} <\min_{1\leq i\leq N}t_{i}$, a contradiction with (\ref{equation: inequality min epsilon2}). Hence, and also using the standard minimax theorem (see, e.g., Theorem 2.10.2 in \cite{zalinescu2002}),
		\begin{align*}
			\min_{\lambda\in\Delta(N)}\max_{\bm t\in \mathcal P}\Bigg\{-t_{0}+\sum_{i=1}^{N}\lambda_{i}t_{i}\Bigg\}= \max_{\bm t\in \mathcal P}\min_{\lambda\in\Delta(N)}\Bigg\{-t_{0}+\sum_{i=1}^{N}\lambda_{i}t_{i}\Bigg\}\leq \frac{\epsilon_{2}}{2}.
		\end{align*}
		Since both $\Delta(N)$ and $\mathcal P$ are compact and convex, we select $\lambda\in \Delta(N)$ that is part of a saddle point and note that $R(E)\coloneq P_{0}(E)-\sum_{i=1}^{N}\lambda_{i} P_{i}(E)\leq \frac{\epsilon_{2}}{2}$ for all $E\in \Sigma_{S}$. Taking complements, we infer that $\abs{R(E)}\leq \frac{\epsilon_{2}}{2}$. Therefore, $\abs{R(E_{1}) - R(E_{2})}\leq \epsilon_{2}$ for all $E_{1},E_{2}\in\Sigma_{S}$. By the Hahn-Jordan decomposition, $R = R^{+}-R^{-}$, where $R^{+}$ and $R^{-}$ are positive measures. In particular $R^{+}(S) = R(E_{1})$ and $R^{-}(S) = - R(E_{2})$ for some $E_{1},E_{2}\in\Sigma_{S}$, and thus $\norm{R}_{1}\leq \epsilon_{2}$.
		
		To show that (ii) implies (i), the pair $(u_{0},\U)$ satisfies $\epsilon_{1}$-Lottery Pareto as a consequence of Proposition \ref{proposition: equivalent forms of approximation} and Theorem \ref{theorem: semistrong without compactness} applied to the subdomain with vN-M utilities induced by lotteries with agreed-upon probabilities. Now assume that, given $E\in\Sigma_{S}$ and $F\in\Sigma_{u}$,  $u_{i}(c^{\ast}Ec_{\ast})\geq u_{i}(c^{\ast}Fc_{\ast})$ for $i=1,\dots, N$. Then $ P_{0}(E)-R(E) =\sum_{i=1}^{N}\lambda_{i}P_{i}(E) \geq \sum_{i=1}^{N}\lambda_{i}P_{i}(F)=P_{0}(F)$, where the last equality follows from the definition of the set $\Sigma_{u}$. By the Hahn Decomposition Theorem, we have that $R(E) = R^{+}(E) - R^{-}(E)$ as in the last paragraph, so we deduce that $R(E)\geq -R^{-}(E)$. Therefore, 
		\begin{align}\label{equation: inequalities P epsilon2}
			P_{0}(E)\geq P_{0}(F) + R(E)\geq P_{0}(F) - R^{-}(E)\geq P_{0}(F) - \frac{\epsilon_{2}}{2}.
		\end{align}
		 If $\frac{\epsilon_{2}}{2}> P_{0}(F)$ then $F_{\tfrac{\epsilon_{2}}{2}} = F$ and thus, for $G = F\setminus F_{\tfrac{\epsilon_{2}}{2}} =\emptyset $, $u_{0}(c^{\ast}Ec_{\ast}) = P_{0}(E)(u_{0}(c^{\ast})-u_{0}(c_{\ast})) + u_{0}(c_{\ast}) \geq   u_{0}(c_{\ast}) =u_{0}(c^{\ast}Gc_{\ast})$. When $\frac{\epsilon_{2}}{2}\leq  P_{0}(F)$ it follows from the right-most inequality in (\ref{equation: inequalities P epsilon2}) that, for $G = F\setminus F_{\tfrac{\epsilon_{2}}{2}}$, $u_{0}(c^{\ast}Ec_{\ast}) = P_{0}(E)(u_{0}(c^{\ast})-u_{0}(c_{\ast})) + u_{0}(c_{\ast}) \geq  \left(P_{0}(F) - \frac{\epsilon_{2}}{2}\right)(u_{0}(c^{\ast})-u_{0}(c_{\ast})) +  u_{0}(c_{\ast}) =u_{0}(c^{\ast}Gc_{\ast})$.


\bibliographystyle{chicago}
\bibliography{mybib.bib}

%


\end{document}